\documentclass[a4paper,twoside,12pt]{article}
\usepackage[authoryear,round]{natbib}
\usepackage{graphicx}
\usepackage{amsmath}
\usepackage{amssymb}
\usepackage{url}
\usepackage{enumitem}       
\usepackage{xr}
\usepackage{color}

\newcommand{\argmax}{\operatornamewithlimits{argmax}}

\newcommand{\Nn}[1]{N_{n}(#1)}
\newcommand{\Dn}[1]{D_{n}(#1)}

\newcommand{\gn}{\hat{g}_{n}}
\newcommand{\g}{g_0}
\newcommand{\G}{\mathbb{G}}

\newcommand{\conv}[1]{\overset{#1}\rightarrow}
\newcommand{\nconv}[1]{\overset{#1}\nrightarrow}

\newcommand{\intinfinf}{\int_{-\infty}^{\infty}}
\newcommand{\intzeroinf}{\int_{0}^{\infty}}
\newcommand{\intainf}[1]{\int_{a}^{\infty}}

\newcommand{\SUM}[3]{\sum_{#1=#2}^{#3}}
\newcommand{\nSUM}[4]{\sum^{#3}_{\substack{#1=#2 \\ (\neq #4)}}}
\newtheorem{theorem}{Theorem}[section]
\newtheorem{lemma}[theorem]{Lemma}


\begin{document}
\thispagestyle{empty}

\noindent
{\LARGE{\bf Feature Sensitive and Automated Curve Registration}}

\bigskip
\noindent
DIBYENDU BHAUMIK\\
{\it Department of Statistics and Information Management, Reserve Bank of India, Mumbai, India}

\noindent
RADHENDUSHKA SRIVASTAVA\\
{\it Department of Mathematics, Indian Institute of Technology, Bombay, India}

\noindent
DEBASIS SENGUPTA\\
{\it Applied Statistics Unit, Indian Statistical Institute, Kolkata, India}

\bigskip
\noindent
{\bf ABSTRACT}
Given two sets of functional data having a common underlying mean function but different degrees of distortion in time measurements, we provide a method of estimating the time transformation necessary to align (or `register') them. We prove that the proposed method is consistent under fairly general conditions. Simulation results show superiority of the performance of the proposed method over two existing methods. The proposed method is illustrated through the analysis of three paleoclimatic data sets.


\bigskip
\setcounter{section}{1}
\baselineskip=18pt
\noindent
{\bf 1. Introduction}\label{sec_intro}\\
Consider functional data arising from observations recorded at a sequence of time points. The task of aligning multiple but similar sets of functional data by possibly nonlinear adjustment to their time scales is often referred to as `registration'. The problem of registration is also important in image and video processing, where multiple dimensions are involved. In the one dimensional case, the need for registration has been felt in broadly two types of problems. For longitudinal growth data, often viewed as a common pattern expressed differently through different individuals with their diverse scales of evolution, the need for registration arises from the quest for the common pattern. In this type of problems, the number of individuals is generally much more than the number of observations per individual. On the other hand, for paleoclimatic data on historic movement of climate variables, the need for registration arises from the notion that the recorded `time' contains estimation error that might come in the way of collation of information from multiple sets of data. In these applications, the number of observations per data set is much more than the number of data sets to be time-aligned or registered. There may even be only two data sets for alignment. In this paper, we address the second type of problems, and focus on registration of one data set with respect to another. While registration may be needed for the purpose of comparing/correlating one variable with another, we consider it only in the context of pooling two data sets on a common variable for the ultimate purpose of better description of its movement over time.

\begin{figure}[t]
 \centering
  \includegraphics{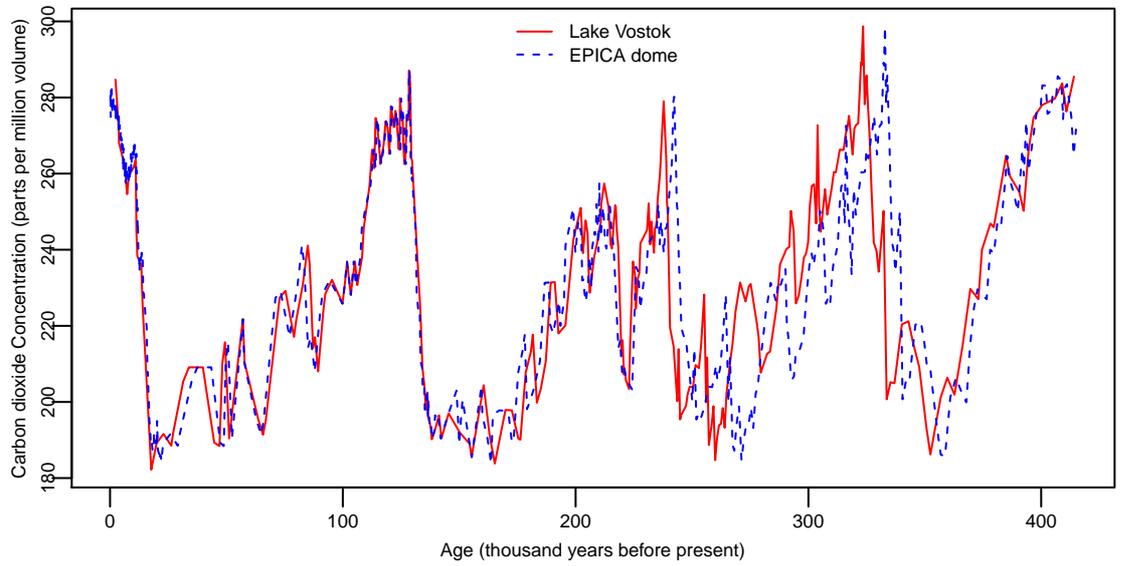}
  \caption{Ice core data on the atmospheric concentration of carbon dioxide}\label{fig:co2}
\end{figure}

As a motivating example, consider the atmospheric concentration of carbon dioxide for the past 400,000 years constructed from two ice cores extracted from two different locations in Antarctica, plotted in Figure~\ref{fig:co2}. One of these data series was collected from the ice sheet over lake Vostok \citep{Petit_et_al_1999}, and the other from EPICA dome \citep{Luthi_et_al_2008}. The peaks in the EPICA dome data (many of which are sharp) precede those in the lake Vostok data during the initial and later parts of the time span, while the reverse happens in the middle part. It suggests the possible existence of a non-linear relationship between the time values of the two data sets.

\begin{center}
Figure~\ref{fig:co2} here.
\end{center}

\pagestyle{myheadings}
\markboth{Feature sensitive automated registration}{Bhaumik et al.}

Sharp peaks and valleys present in functional data sets may be viewed as either a help or a hindrance for registration. One may hold the view that paying too much attention to these details may cause distortion, as observation times in either of the data sets may miss the actual peaks of the underlying continuous time phenomenon. In that case, the method of registration should not utilize these features as such. Several methods of this kind are available: continuous monotone registration \citep{Ramsay_Li_1998}, dynamic time warping \citep{Wang_Gasser_1997, Wang_Gasser_1999}, registration by local regression \citep{Kneip_et_al_2000}, maximum likelihood registration through parametric modeling of the time transformation \citep{Ronn_2001, Gervini_Gasser_2005}, self-modelling warping \citep{Gervini_Gasser_2004}, shape invariant model based registration \citep{Brumback_Lindstrom_2004}, functional convex synchronization model based registration \citep{Liu_Muller_2004}, pair-wise curve synchronization \citep{Tang_Muller_2008}, and functional principal component based registration \citep{Silverman_1995, Kneip_Ramsay_2008}.

One may also hold the view that sharp peaks and valleys in the data are characteristics of the underlying common function, and in that case it may be unwise to ignore the information contained in them. Making use of such information may yield better registration. Smoothing based methods may be unable to take full advantage of this information, as smoothing blurs these features.

It is possible to use only the sharp peaks and valleys in the data series, often identified as `landmarks', for the purpose of registration. A simplistic method would be to identify specific landmarks of one data set that correspond to the landmarks of the other data set, and use a piecewise linear time transformation that permits the alignment of the two sets of landmarks. \cite{Ramsay_Li_1998} suggested that such a crude registration may be further refined through their method. The other methods mentioned above may also be used for refinement. Manual identification of some matching landmarks in the two data sets (after preliminary smoothing) have been used as input to the methods proposed by \cite{Kneip_Gasser_1992} and \cite{Kneip_Engel_1995}. The method of \cite{Kneip_Gasser_1992} leads to registration under the constraint that the identified landmarks should match. The identification of matching landmarks is generally done manually. This is a disadvantage of landmark-assisted registration.

\cite{Bigot_2006} proposed an automated landmark-based method, which provides registration through the following steps: (a)~identification of significant landmarks in each of the two data sets, (b)~establishing possible correspondence between these landmarks, and (c)~estimation of the time transformation function through a standard method of nonparametric smoothing/regression, by using the pairs of matched times of landmarks as input. The performance of this multiple-step method may
be limited by the non-use of information other than landmarks and possible accumulation of estimation errors at different steps.

\cite{James_2007} proposed a new method based on matching of `functional moments', intended to capture landmarks or local features. This method, as well as the functional principal components used by \cite{Silverman_1995} and \cite{Kneip_Ramsay_2008}, requires a considerable number of data sets (which is natural in longitudinal growth data) for estimation of the population characteristics, and are not usable for applications where only a pair of data sets need to be registered.

With the objective of registering one functional data set with respect to another, we model the observations of the two data sets as
\begin{eqnarray*}
y_1(t)&=&m(t)+\epsilon_1(t) \\
y_2(t)&=&m(g(t))+\epsilon_2(t)
\end{eqnarray*}
where $m$ is the underlying mean function, $g$ is the requisite time transformation, and $\epsilon_1$ and $\epsilon_2$ are random error terms. The task of registration is essentially that of estimating $g$ that aligns the two sets of data. Note that the above model would not be suitable for data sets representing observations on different variables (e.g., paleoclimatic data sets on temperature and carbon dioxide). On the other hand, any generalization of the above model for applicability to different variables would involve additional estimation, which is redundant when the observed variables are in fact identical. Thus, such a generalization can only be achieved at the cost of inefficient utilization of the available information.

In Section~2, we propose a new estimator of the time transformation function $g$, by maximizing a `measure of alignment' of two functional data sets. The maximization is done over an appropriate class of transformations. This measure of alignment is designed to capture the information contained in the entire set of data, including locations of sharp variation. We show that it possesses some desired characteristics. The method is automated as it does not require manual identification of landmarks. Identifiability of $g$ with respect to the chosen model, under appropriate conditions, is proved in Section~3.

Many of the existing methods of estimation of the time transformation have been proposed without establishing their consistency.
\cite{Ronn_2001} and \cite{Gervini_Gasser_2005} have proved the consistency of their methods when the number of subjects goes to infinity. Following \cite{Kneip_Engel_1995}, \cite{Wang_Gasser_1999}, and \cite{Gervini_Gasser_2004}, we have established the consistency of our estimator as the numbers of observations in the two data sets go to infinity. This result, reported in Section~3, holds when the time transformation is chosen from a class of functions satisfying some general conditions.

Results of a simulation study, to demonstrate the performance of the estimator chosen from a particular class, are reported in Section~4. The method is illustrated in Section~5 through the analysis of several real data sets. Some concluding remarks are provided in Section~6.

\setcounter{section}{2}
\bigskip
\noindent
{\bf 2. Model and Methodology}\label{sec_mod_and_met}\\
Let $\{(t_1, Y_{t_1}),\ldots,(t_{n_1}, Y_{t_{n_1}})\}$ and $\{(s_1,Y_{s_1}'),\ldots,(s_{n_2},Y_{s_{n_2}}')\}$ be two sets of functional data, arising from the model
\begin{eqnarray}\label{model1}
Y_{t_i} &=& m(t_i)+\epsilon_1(t_i)\qquad\qquad\ \  i=1,\ldots,\,{n_1}\notag\\
Y_{s_j}'&=&m(g_0(s_j))+\epsilon_2(s_j),\qquad j=1,\ldots,\,{n_2}
\end{eqnarray}
where $m$ is an underlying location function that is continuous, and $g_0$ is an unknown time transformation function, which is continuous and strictly increasing. The terms $\epsilon_1$ and $\epsilon_2$ represent additive random measurement errors, which have mean zero.

Now, let us define, for any given continuous and strictly increasing transformation function $g$, the functional
\begin{equation}\label{cr1}
L_n(g)=
\frac
{
\displaystyle\frac{1}{n_1 n_2}
\displaystyle\sum\limits_{i=1}^{n_1}
\displaystyle\sum\limits_{j=1}^{n_2}
\frac{1}{h_1}K_1\left(\frac{{t_i} - g({s_j})}{h_1}\right)
\frac{1}{h_2}K_2\left(\frac{{Y_{t_i}} - Y_{s_j}'}{h_2}\right)
}
{
\displaystyle\frac{1}{n_1 n_2}\displaystyle\sum\limits_{i=1}^{n_1}
\displaystyle\sum\limits_{j=1}^{n_2}
\frac{1}{h_1}K_1\left(\frac{{t_i} - g(s_j)}{h_1}\right)},
\end{equation}
where $n=n_1+n_2$, $K_1$ and $K_2$ are kernel functions that are probability densities, and $h_1$ and  $h_2$ are the corresponding bandwidths. The above functional can be interpreted as a weighted sum of the terms $\frac{1}{h_2}K_2\left(\frac{{Y_{t_i}} - Y_{s_j}'}{h_2}\right)$ for $i=1,\ldots,n_1$ and $j=1,\ldots,n_2$. The weights depend on~$g$. Note that when $g=g_0$, for every pair of $i$ and $j$ such that $g(s_j)$ is close to $t_i$, the continuity of $m$ ensures that $Y_{t_i} - Y_{s_j}'$ is expected to be small. Therefore, for $g=g_0$, large values of $\frac{1}{h_2}K_2\left(\frac{Y_{t_i} - Y_{s_j}'}{h_2}\right)$ are expected to occur together with large values of their weights. This may not be the case when $g\neq g_0$. In Section~3, we show that under some general conditions, the probability limit of $L_n(g)$ attains its maximum value if and only if $g=g_0$.

Thus, $L_n(g)$ may be interpreted as a measure of alignment.

Let us now examine the roles of the bandwidth parameters $h_1$ and $h_2$ in the above measure. A small value of $h_1$ makes the weight for a given $i$ and $j$ nearly equal to zero, unless $t_i$ is very close to $g(s_j)$. Thus, only a few weights can be substantial. When $h_1$ is large, weights can be substantial for more combinations of $i$ and $j$. Thus, $h_1$ controls the effective number of weights in the weighted sum in \eqref{cr1}. On the other hand, $h_2$ controls the penalty for discrepancies between $Y_{t_i}$ and $Y_{s_j}'$. A very large value of $h_2$ might make $L_n(g)$ insensitive to changes in $g$, as there would not be enough penalty for mismatch between $Y_{t_i}$ and $Y_{s_j}'$. A very small value of $h_2$ would make $L_n(g)$ unstable, as $\frac{1}{h_2}K_2\left(\frac{{Y_{t_i}} - Y_{s_j}'}{h_2}\right)$ would be nearly zero for most of the combinations of $i$ and $j$.

We define the proposed estimator of the function $g_0$ as
\begin{equation}\label{cr2}
\hat{g}_n=\arg\max_{g\in\G} L_n(g),
\end{equation}
where $L_n(g)$ is as defined in (\ref{cr1}) and $\G$ is a suitable class of continuous and strictly increasing functions that includes the true transformation function~$g_0$.

As for choices of the bandwidths, one can select $h_1$ as a fraction of the range of time in either sample and $h_2$ as a fraction of the combined range of the observed variable. Some guidelines are given in Section~4.

Corresponding peaks in two sets of functional data are sometimes identified manually as matching landmarks. In our case, the objective function $L_n(g)$ automatically rewards candidate transformation functions that map peaks of one data set into the corresponding peaks of the other. On the other hand, if a peak is missing from one of the data sets, then it does not penalize the `correct' transformation any more than a similar alternative candidate (i.e., a marginally different transformation). Therefore, the estimator $\hat{g}_n$ should be able to utilize the landmarks automatically for registration.

\setcounter{section}{3}
\bigskip
\noindent
{\bf 3. Consistency}\label{sec_the_res}\\
Let the errors $\{\epsilon_1(t_i);\, i=1,\ldots n_1\}$ and $\{\epsilon_2(s_j);\, j=1,\ldots n_2\}$ and the time points $\{t_i;\, i=1,\ldots n_1\}$ and $\{s_j;\, j=1,\ldots n_2\}$ be mutually independent sets of samples from the probability density functions $f_{\epsilon_1} $, $f_{\epsilon_2} $, $f_1 $, and $f_2 $ having supports over $[-\infty,\infty]$, $[-\infty,\infty]$, $[a,b]$ (for $0\le a<b$), and $[c,d]$ (for $0\le c<d$) respectively.

We assume that $m$ in model~(\ref{model1}) is a continuous function defined over $[0,\infty)$. We also presume that it is not flat or it does not fluctuate too much. In order to ensure this, we formally stipulate that for any given interval $[p,q]$, the inverse image (with respect to $m$) of any point on $m([p,q])$ has a finite intersection with $[p,q]$, i.e., for every $y\in\left[\min_{t\in[p,q]}m(t),\max_{t\in[p,q]}m(t)\right]$, the set $\left\{t:\ m(t)=y,\ t\in[p,q]\right\}$ has a finite number of elements.

Estimation of $g_0$ in model~(\ref{model1}) makes sense only if $g_0([c, d])$ has a substantial overlap with $[a, b]$. We, therefore, assume that $g_0([c, d])\cap [a, b]$ includes a non-empty open interval. We also need to ensure that there is no ambiguity about $g_0$ in model~\eqref{model1}. Let $\G_0$ be the class of all strictly increasing and continuous functions $g$ defined over $[c, d]$ such that the set $S_g=g^{-1}([a, b]) \cap [c, d]$ contains a non-empty open interval and that $g$ agrees with $g_0$ at least at one point in $S_{g_0} \cap S_g$. By construction, $g_0\in\G_0$. Our first result establishes the identifiability of $g_0$ (within $\G_0$) with respect to model~(\ref{model1}).

\begin{theorem}\label{m_thm0}
Let $m$, $g_0$, and $\G_0$ be as described above. If $g\in\G_0$ is such that $m(g(s))=m(g_0(s))$ for all $s\in S_g\cap S_{g_0}$, then $S_g=S_{g_0}$ and $g(s)=g_0(s)$ for all $s\in S_{g_0}$.
\end{theorem}

The consistency of $\hat{g}_n$ needs to be established as the sample size in both the data sets go to infinity \citep{Kneip_Engel_1995,Wang_Gasser_1999,Gervini_Gasser_2004}.
As a first step, we establish the point-wise convergence of the functionals $L_n$ on $\G_0$ after making the following assumptions.

\medskip\noindent
{\bf Assumption 1.} The densities $f_{\epsilon_1}$, $f_{\epsilon_2}$, $f_1$, and $f_2$ are continuous and bounded; $f_{\epsilon_1}$ and $f_{\epsilon_2}$ are symmetric about zero and are strictly unimodal at zero; $f_1$ and $f_2$ are positive over the interior of their supports.

\medskip\noindent
{\bf Assumption 2.} The kernels $K_1$ and $K_2$ are continuous and bounded probability density functions defined over
the real line.

\noindent
This condition is satisfied by all the popular kernels viz. Uniform, Triangular, Epanechnikov, Biweight, Gaussian, and so on.

\medskip\noindent
{\bf Assumption 3.} The sample sizes $n_1$ and $n_2$ are such that $n_1/n\rightarrow\xi$ for some $\xi\in(0,1)$, as $n\rightarrow\infty$.

\medskip\noindent
{\bf Assumption 4.} The bandwidths are such that $h_i\rightarrow0$ and $n_ih_i\rightarrow\infty$ as $n\rightarrow\infty$, $i=1,2$.

\medskip
\noindent
\begin{theorem}\label{m_thm1}
Let $m$, $g_0$, and $\G_0$ be as described at the beginning of this section. Then, under Assumptions 1--4, for any function $g\in\G_0$, as $n\rightarrow\infty$, $L_n(g)\conv{P}L(g)$, where
\begin{eqnarray}\label{exp_C}
L(g)=\frac{\intinfinf\intzeroinf f_1(g(y))f_2(y)f_{\epsilon_1}(v-m(g(y))+m(g_0(y)))f_{\epsilon_2}(v) dydv}{\intzeroinf f_1(g(y))f_2(y)dy}.
\end{eqnarray}
\end{theorem}

\medskip\noindent
We now show that this limiting functional is maximized only by the correct transformation function.

\medskip\noindent
\begin{theorem}\label{m_thm1.5}
Suppose $m$, $g_0$, and $\G_0$ are as described at the beginning of this section. Then, under Assumption 1,
\begin{enumerate}
\item[(a)] $L(g)\le L(g_0)$ for all $g\in\G_0$,
\item[(b)] If $L(g)= L(g_0)$ for some $g\in\G_0$,
then $g=g_0$ over $S_{g_0}$.
\end{enumerate}
\end{theorem}

\noindent
The next step is to establish the uniform convergence of $L_n$, for which we need a stronger condition on $\G_0$ that enforces compactness.
Let us define a metric on $\G_0$ viz., $d(g_1, g_2)=\sup_{x\in[c,d]}|g_1(x)-g_2(x)|=\|g_1-g_2\|;\; g_1, g_2 \in G_0$.

\medskip \noindent{\bf Assumption 5.} The class $\G$ in~(\ref{cr2}) is a compact subset of $\G_0$ in the metric space $(\G_0, d)$ and it includes $g_0$.

\medskip\noindent
An example of $\G$ that satisfies Assumption~5 is the subset of functions $g$ of $G_0$ with bounded slope.

\medskip\noindent
{\bf Assumption 2A.} The kernels $K_1$ and $K_2$ are bounded away from zero on a given closed interval and have bounded first order derivatives.

\noindent
The Gaussian kernel satisfies the above condition.

\begin{theorem}\label{m_thm2}
Let $m$, $g_0$, and $\G_0$ be as described at the beginning of this section. Then under the Assumptions 1--5, and 2A, as $n\rightarrow\infty$, $$\sup_{g\in\G}\left|L_n(g)-L(g)\right|\conv{P}0.$$
\end{theorem}

\noindent
We now establish that the sequence of maximum values of the functionals $L_n$ converges to the value of $L$ at its maximizer, $g_0$.

\begin{theorem}\label{m_thm3}
Let $m$, $g_0$, and $\G_0$ be as described at the beginning of this section. Then under the Assumptions 1--5, and 2A,  $L_n(\hat{g}_n)\conv{P} L(g_0)$ as $n\rightarrow\infty$.
\end{theorem}

\noindent
Finally we establish the consistency of our estimator.

\begin{theorem}\label{m_thm4}
Let $m$, $g_0$, and $\G_0$ be as described at the beginning of this section. Then under the Assumptions 1--5, and 2A, $\hat{g}_n\conv{P} g_0$ as $n\rightarrow\infty$.
\end{theorem}

\setcounter{section}{4}
\bigskip
\noindent
{\bf 4. Simulation of performance}\label{sec:sim_app}\\
\setcounter{subsection}{1}
\medskip
\noindent
{\sl 4.1.Methods compared}\\
Even though we have proposed a general class of estimators in Section~2 and established their consistency in Section~3, we need to focus on a specific member of that class in order to simulate the performance. We chose $\G$ as the vector space generated by linear B-spline basis functions with equidistant knot points over $[a, b]$. The method of Steepest Ascent was used to maximize~(\ref{cr1}). We opted for standard normal densities for $K_1$ and $K_2$. We chose $h_1$ as 5 per cent of the range of time in either data set, and $h_2$ as $10$ per cent of the range of combined $Y$-values of both the data sets. In order to provide an initial iterate to steepest ascent, we programmatically mapped significant peaks and valleys of the two data sets. The optimization algorithm, however, produced reasonably good results (not reported here) even when the identity map was provided as initial iterate.

We carried out simulations to compare the performance of the above implementation of the proposed method with two other methods.
\begin{enumerate}
\item	The first method was continuous monotone registration. We used its implementation in the R-function {\small\tt register.fd} in the package {\small\tt fda}. We retained default values of the order of the polynomial splines~(4) and the roughness penalty~(2). Following \cite{Ramsay_Li_1998}, we selected the value of the smoothing parameter $\lambda$ as $10^{-3}$.
\item	The second method is self-modelling registration. We used the matlab codes provided by Daniel Gervini on his web site (\url{https://pantherfile.uwm.edu/gervini/www/}). We chose default values for the number of random starts (20) and the order of splines (3). Following \cite{Gervini_Gasser_2004} other parameters of the method viz. the numbers of components ($q$) and the number of basis functions ($p$) were selected by using the cross-validation algorithm.
\end{enumerate}
These methods were selected for comparison mostly on account of availability of codes and applicability to the data at hand. The automated method of \cite{Bigot_2006} was not chosen for comparison, as his code can only handle sample sizes that are powers of~2.

\setcounter{subsection}{2}
\bigskip
\noindent
{\sl 4.2. Simulation Design}\\
The choice of data for simulation was motivated by the paleoclimatic data set that exhibits sharp changes. We chose lake Vostok data on atmospheric concentration of carbon dioxide (with $283$ data points, range of time-values $I=[2.3,414.1]$ and s.\ d.\ of $Y$-values as $s=28.7$) described in Section~1, as the base for our simulation exercises. We conducted Monte Carlo simulations in four different scenarios under model~(\ref{model1}) (with $n_1=n_2=250$) as described below:

\begin{description}
\item[Scenario 1.] Here we randomly selected a sub-set $D$ of $250$ data points from the base set. The function $m$ was obtained by linear interpolation from those selected points. In every simulation run, we constructed the pair of data sets as follows.
\begin{enumerate}[label=(\alph*)]
\item	Time values of both the data sets were kept same as those of $D$.
\item	The time transformation $g_0$ in~(\ref{model1}), applied on the time values of the second data set, was chosen as the linear spline with six equidistant knots over the range of time-values of $D$ (see graph in Figure~\ref{fig:sim_g0}(i) in Supplemental Materials), i.e.,
\begin{eqnarray}
     g_0(t)&=&-0.379 + 1.05t - 0.209(t-85.7)_{+} \nonumber\\
     &&+\, 0.409(t-167.8)_{+} - 0.609(t{-}249.9)_{+} \nonumber\\
     &&+\, 0.809(t{-}332.0)_{+}\:\mbox{, for {\em t}}\in [3.6,414.1]
    \label{sim_g_1}
\end{eqnarray}

where
\begin{equation*}
  u_+ = \left\{
  \begin{array}{rl}
  u & \text{if } u > 0,\\
  0 & \text{if } u \le 0.
  \end{array} \right.
\end{equation*}
\item	Additive errors were generated afresh from the normal distribution (mean=$0$, s.d.=$0.05s$) for both the data sets separately.
\end{enumerate}

\item[Scenario 2.] Here, $m$ was the function obtained by linearly interpolating between the data points of the base data. The $250$ time points as well as the additive errors for each data set were generated afresh for each simulation run. In this case, the distributions $f_1$ and $f_2$ described in Section~3 were uniform over $I$ (see the beginning of this subsection), while the distributions $f_{\epsilon_1} $, $f_{\epsilon_2}$ were normal (mean=$0$, s.d.=$0.05s$). The function $g_0$ was as in~(\ref{sim_g_1}).

\item[Scenario 3.] This set-up was the same as in Scenario~1, except that $g_0$ was chosen as the identity function plus a periodic function viz.,
\begin{eqnarray}
g_0(t)=t+0.05t\sin(3\pi t/b)\:\mbox{for}\: t \in [a, b].
\label{sim_g_2}
\end{eqnarray}
The graph of this function is shown in Figure~\ref{fig:sim_g0}(ii) in Supplemental Materials.

\item[Scenario 4.] This set-up was the same as in scenario 2, except that $g_0$ was as in~(\ref{sim_g_2}).
\end{description}
Scenario 1 offers an opportunity to assess the performance of the methods when $\G$ includes $g_0$. We chose 21 equidistant knots in our search space of B-splines, so that $g_0$ with six equidistant knots becomes a special case. In Scenarios 2, 3, and 4, the chosen number of knots was $20$, which means that $g_0$ is not included in the search space. For continuous monotone registration, the number of knots chosen for each scenario was the same as that of the proposed method.

The matlab programs for self-modelling registration require values of the functions to be registered at a common set of time points, which is met only in scenario $1$ and $3$. Therefore, simulation results for this method are reported only for these two scenarios.

The performance of the estimators of $g_0$ were studied in terms of (a)~point-wise bias, (b)~point-wise standard deviation, (c)~point-wise mean squared error (MSE), and (d)~average of the integrated mean square error (IMSE) normalized by the squared norm of the true function, defined for each simulation run as $$\frac{\frac1{S}\sum_{j=1}^{S}\int_a^b(\hat{g}_j(t)-g_0(t))^2dt}{\int_a^b g_0^2(t)dt},$$ where $S$ is the number of independent runs of the simulation and $\hat{g}_j$ is the estimate of $g_0$ at the $j$th run. We used Simpson's rule to evaluate these definite integrals.

\begin{figure}[t]
 \centering
  \includegraphics{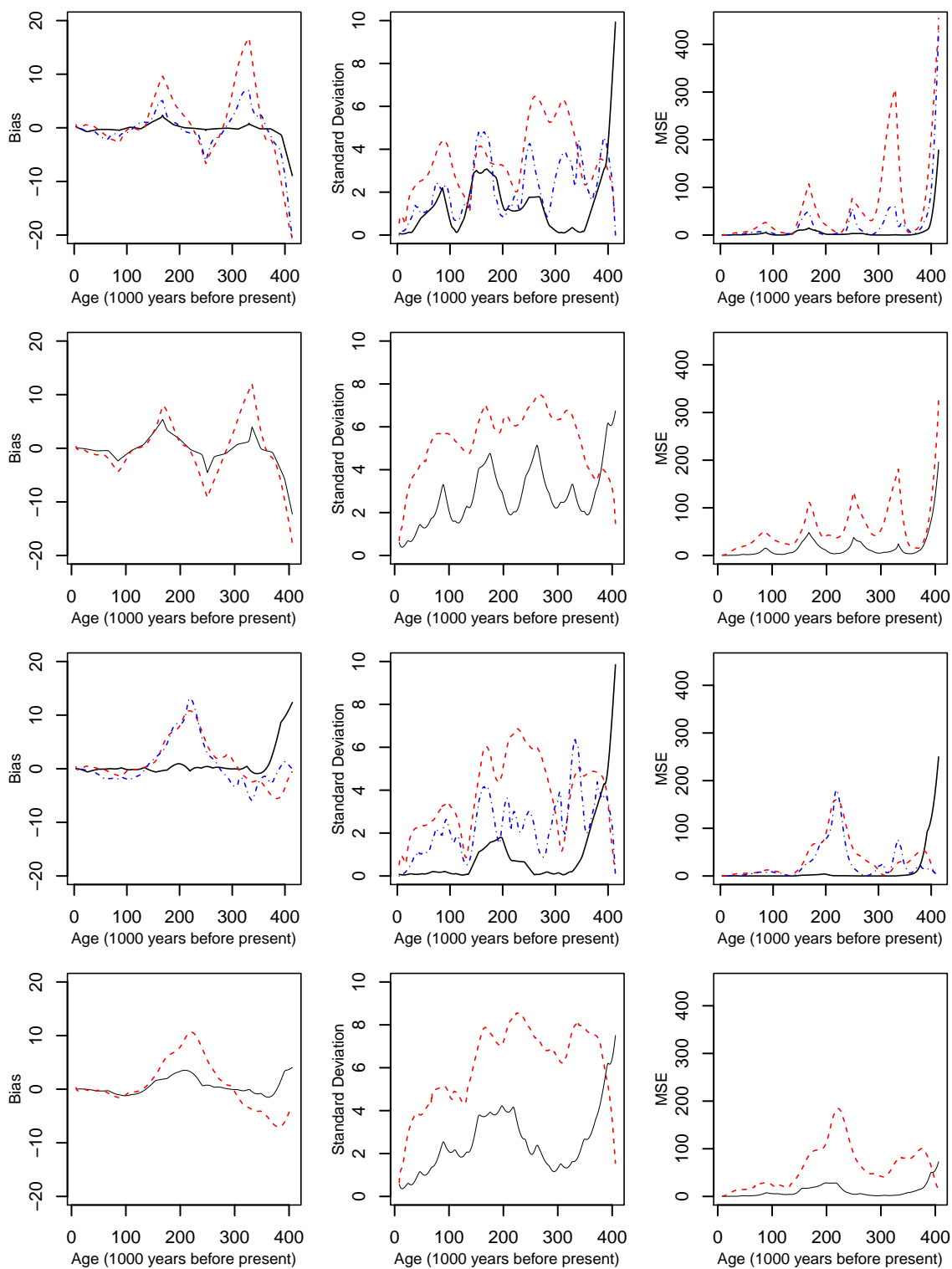}
  \caption{Point-wise bias, standard deviation, and MSE of the estimators of $g_0$ by continuous monotone registration (broken line), self-modelling registration (dotdash) and the proposed method (solid) under scenario $1$ (top row), scenario $2$ (second), scenario $3$ (third), and scenario $4$ (bottom)}\label{fig:bias_sd_mse}
\end{figure}

\setcounter{subsection}{2}
\medskip
\noindent
{\sl 4.3. Results}\\
The results of the simulations based on $1000$ independent runs for each of the scenarios, are shown in Figure~\ref{fig:bias_sd_mse}. In scenarios~$2$ and $4$, where time-points change from one run to another, the R-function {\small\tt register.fd} did not produce results for $240$ simulation runs. Therefore, results for this estimator corresponding to these scenarios are based on $760$ runs.

\begin{center}
Figure~\ref{fig:bias_sd_mse} here.
\end{center}

\begin{center}
Table~\ref{tab:imse} here.
\end{center}

It is observed that, the proposed estimator has, in general, smaller bias, standard deviation, and MSE as compared to continuous monotone registration, and self-modelling registration. However, towards the right end of the time scale in Scenarios~3 and~4, the proposed method exhibits both higher standard deviation, and MSE as seen in Figure~\ref{fig:bias_sd_mse}. In this context, it may be mentioned that Kernel based methods are known to perform poorly near the fringes of the data. It is observed in Table~\ref{tab:imse} that the proposed method had uniformly smaller average normalised IMSE for all the scenarios.

\setcounter{section}{5}
\bigskip
\noindent
{\bf 5.	Analysis of ice core data}\\
Here we considered paleoclimatic data on the atmospheric concentration of (i) carbon dioxide, and (ii) methane \citep{Petit_et_al_1999, Loulergue_et_al_2008} as determined from air-bubbles trapped in ice cores collected over Lake Vostok and at EPICA Dome of Antarctica, and (iii) average annual temperature deviations \citep{Petit_et_al_1999, Valerie_2007}, which were reconstructed from deuterium contents at various depths of ice cores obtained at these two sites. Table~\ref{tab:des_stat} gives some descriptive statistics of the data sets.

\begin{center}
Table~\ref{tab:des_stat} here.
\end{center}

We chose to align the data set from EPICA dome with that from lake Vostok using two registration methods viz.\ the proposed method, and continuous monotone registration. We could not apply the method of self-modelling registration for the above data, since its matlab implementation requires the nominal observation times in the two data sets to coincide.

The number of knots used for both the methods compared here were $30$, $30$, and $20$ for carbon dioxide, methane, and temperature deviation data respectively, while remaining parameters were chosen as in the previous section.

\begin{figure}[t]
  \includegraphics{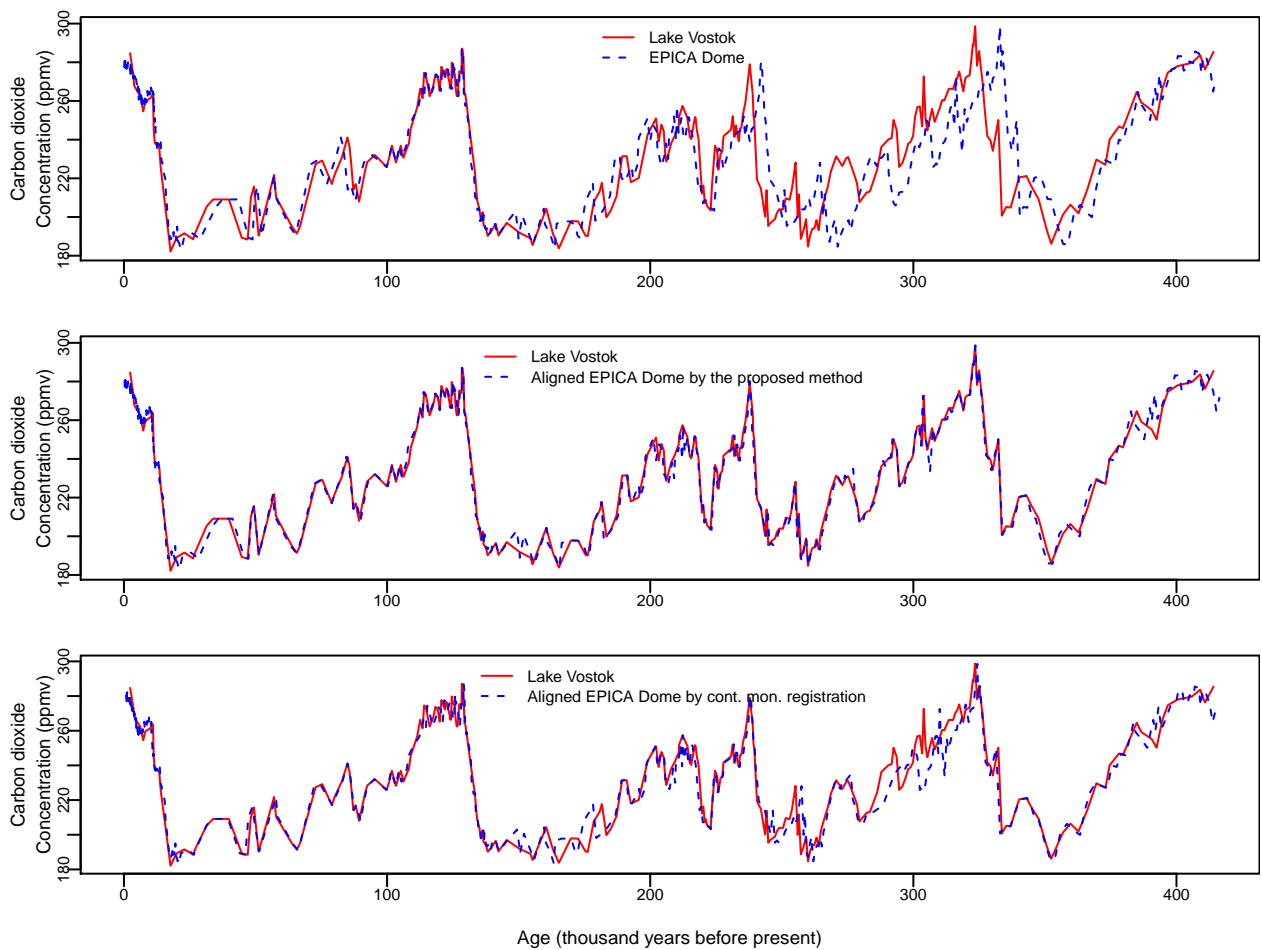}
  \caption{Alignment of data sets on atmospheric concentration of carbon dioxide}\label{fig:co2_aligned}
\end{figure}

\begin{figure}[t]
  \includegraphics{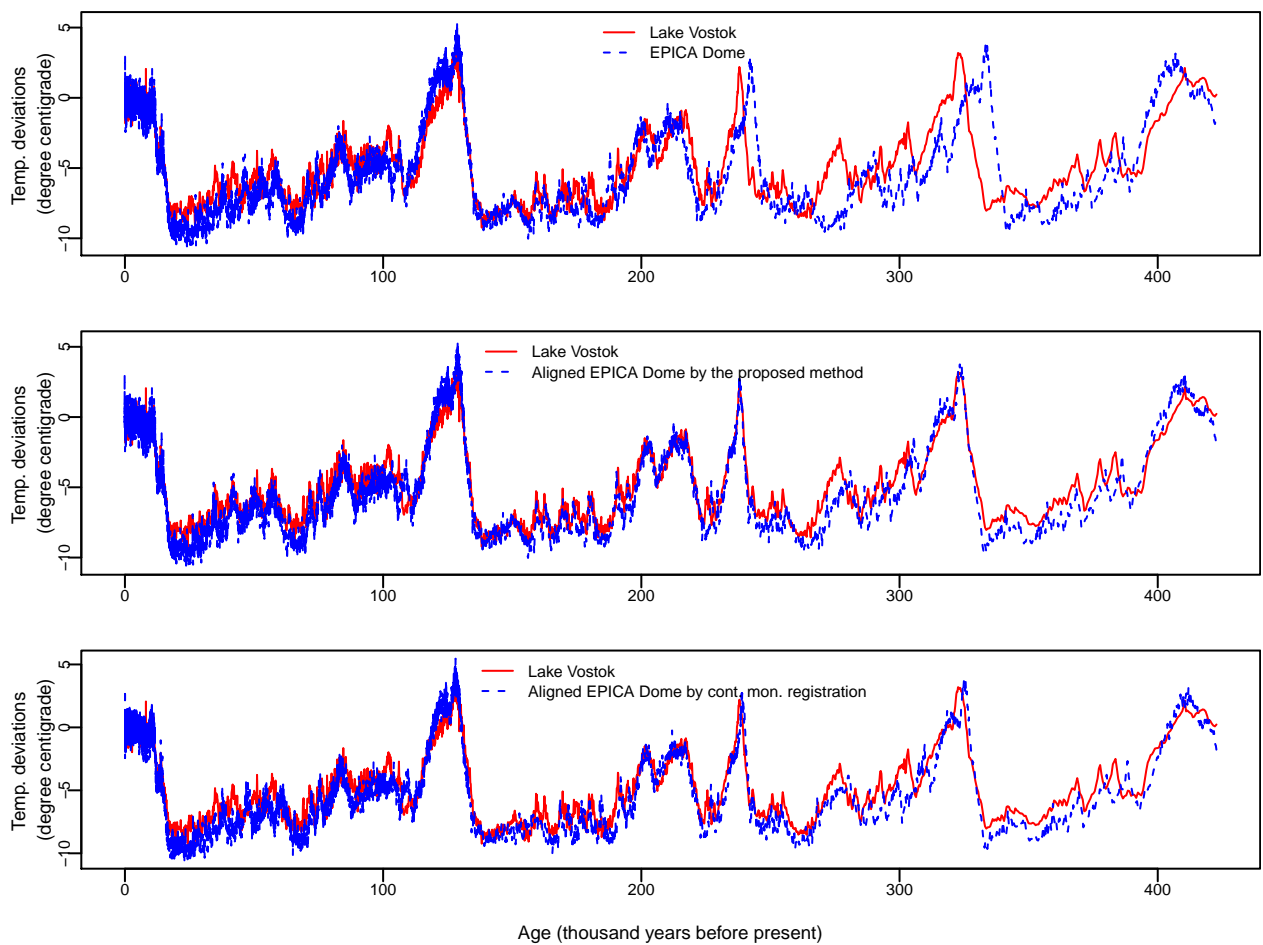}
  \caption{Alignment of data sets on average temperature deviations}\label{fig:temp_aligned}
\end{figure}

\begin{figure}[t]
  \includegraphics{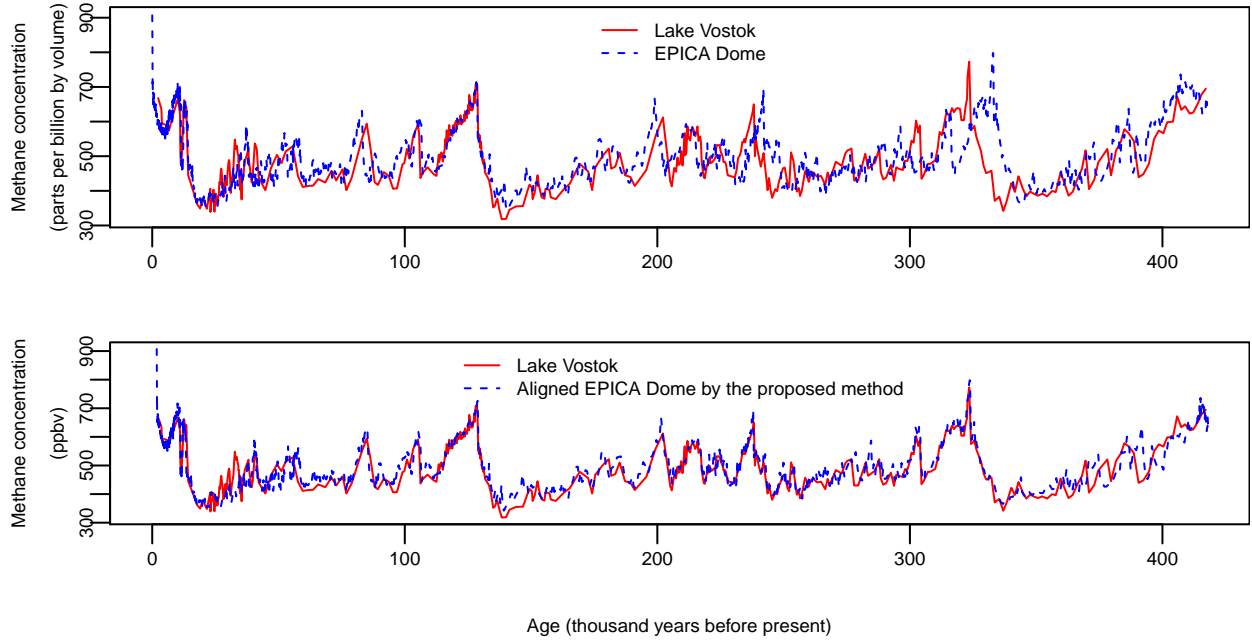}
  \caption{Alignment of data sets on atmospheric concentration of methane}\label{fig:ch4_aligned}
\end{figure}

Alignments of the carbon dioxide, temperature, and methane data sets are shown in Figures~\ref{fig:co2_aligned},~\ref{fig:temp_aligned}, and~\ref{fig:ch4_aligned} respectively. The R-function {\small\tt register.fd} could not produce output for the methane data. The estimates of the function $g_0$ for the three pairs of data sets, produced by the methods are plotted in Figure~\ref{fig:est_g0} of Supplemental Materials.

\begin{center}
Figure~\ref{fig:co2_aligned} here.
\end{center}

\begin{center}
Figure~\ref{fig:temp_aligned} here.
\end{center}

\begin{center}
Figure~\ref{fig:ch4_aligned} here.
\end{center}

\begin{center}
Table~\ref{tab:sq_diff} here.
\end{center}

The average squared distance between the interpolated curves before and after registration by the two methods, computed over a uniform grid of size 1000 over the common time range, are reported in Table~\ref{tab:sq_diff}. It is observed that the proposed method produced shorter distance between the registered curves as compared to continuous monotone registration.

\bigskip
\setcounter{section}{5}
\noindent
{\bf 5. Discussion}\label{sec:dis}\\
In this paper, we have proposed a new method of registration of one functional data set with another, by optimizing an empirical measure of alignment. If there are sharp variations in the data, the proposed method is able to utilize them, without requiring prior identification of landmarks. Since the method does not use any pre-smoothing, it does not suffer from any loss of information that might occur due to smoothing. On the other hand, the measure of alignment (2) ensures that the proposed method makes use of the main strength of smoothing, namely pooling of information from neighbouring observations.

The present implementation of the method, in the form of an R code, is available from the authors on request. This implementation permits registration of data sets with possibly unequal, irregularly spaced and large number of samples. This implementation is based on some specific choices, e.g., use of the class of B-splines with uniformly spaced knot points as candidate time transformation functions, and steepest ascent for optimization. However, none of these choices is necessary in the general set-up used for proving the consistency of the proposed class of estimators.

There are indeed some limitations of the proposed approach. The alignment provided by the proposed method is likely to change if the data sets for registration are interchanged. When there are many sharp changes in the data, the iterative algorithm may run into spurious local maxima, particularly when the bandwidth parameter $h_1$ in (2) is small. Such problems may be mitigated by replacing steepest ascent with a probabilistic search algorithm such as simulated annealing. On the other hand, when the data sets do not have sharp changes, the proposed method may not be very sensitive to alterations in the transformation function. The guideline on the choice of bandwidths (particularly $h_2$) given here may not be applicable for those problems. The proposed method would be completely unsuitable for longitudinal data with many individuals but relatively fewer observations per individual.

When there are more than two data sets to be registered, the proposed method has to be used multiple times on pairs of data, possibly after identifying one of the data sets as reference for registration. This reference data set may also be selected on a trial basis, and the candidate leading to the best overall alignment may be selected as reference data set at the end.

The proposed method can be used as a tool for structural averaging (which amounts to estimating the function $m$ in model (1)). Large sample properties of the resulting estimator of $m$ and its performance in relation to competing methods need to be studied in future.

Registration of two sets of functional data on different variables (e.g., paleoclimatic data on temperature and carbon dioxide) is sometimes needed for the purpose of studying the relationship between them. The method presented here is not readily applicable to this problem. However, an adaptation may be possible by inserting an unknown amplitude parameter in one of the two equations of model (1). This problem is another possible direction of future work in this area.

\bigskip
\noindent
{\bf Acknowledgement}\\
The work of Radhendushka Srivastava is partially supported by INSPIRE research grant from DST of Government of India and a research seed grant from the Indian Institute of Technology, Bombay.

\bigskip
\noindent
{\bf Supporting information} \\
Additional information for this article is available online.
\begin{description}
\item[Figure S1.] $g_0$ functions for simulations
\item[Figure S2.] Estimates of $g_0$ for paleoclimatic data sets
\end{description}

\small
\bibliographystyle{natbib}
\bibliography{mybib}
\medskip
\noindent
Dibyendu Bhaumik, Department of Statistics and Information Management, Reserve Bank of India, C9, 6th Floor, Bandra-Kurla Complex, Bandra (East), Mumbai - 400051, India.\\
E-mail: dbhaumik@rbi.org.in

\appendix
\section{Appendix: Proof of theoretical results}
\medskip
\noindent
{\bf Proof of Theorem \ref{m_thm0}}
Let $g\in\G_0$ be such that $g(\alpha)= g_0(\alpha)$ for some $\alpha\in S_{g_0}\cap S_g$, $m(g(s)) = m(g_0(s))$ for all $s\in S_{g_0}\cap S_g$, and yet $g(s)\ne g_0(s)$ for some $s\in S_{g_0}\cap S_g$.  We can presume, without loss of generality, that $g(s)<g_0(s)$. Let us assume, for now, $\alpha<s$. Then the set $$\{t:\ m(t)=m(g_0(s))\}\cap[g_0(\alpha), g_0(s)]$$ has at least two elements, $g(s)$ and $g_0(s)$. Let $t^{(1)}<\cdots<t^{(k)}$ be the ordered elements of this set. Clearly, $g_0(s)=t^{(k)}$. Let $g(s)=t^{(i)}$ for some $i<k$.

In order that the functions $m(g(u))-m(g(s))$ and $m(g_0(u))-m(g_0(s))$ coincide for all $u\in[\alpha,s]$, these functions should have exactly the same number of zero crossings over this interval. However, from what we have already observed, the first function has exactly $i$ zero-crossings, while the second function has exactly $k$ zero-crossings, and $i<k$. Therefore, the two functions must differ somewhere on $[\alpha,s]$. Similarly, if $\alpha > s$, the set,
$$\{t:\ m(t)=m(g(s))\}\cap [g(s), g(\alpha)]$$
has at least two elements viz., $g(s)$ and $g_0(s)$. If $s^{(1)}<\cdots <s^{(l)}$ be the ordered elements of the set, then $g(s)=s^{(1)}$ and $g_0(s)=s^{(j)}$ for some $j>1$. Following similar arguments, the  two functions above, which have been presumed to coincide for all $u\in[s, \alpha]$, must differ somewhere on $[s, \alpha]$ as the first function has exactly $l$ zero-crossings, while the second has exactly $l-j+1$.

This contradicts the presumption that $g(s)\ne g_0(s)$ for some $s\in S_g\cap S_{g_0}$. Thus, $g=g_0$ over $S_g\cap S_{g_0}$.

The continuity of $g$ and $g_0$, together with their equality over $S_g\cap S_{g_0}$, implies that $S_g=S_{g_0}$. This completes the proof.$\hfill\Box$

\bigskip
\noindent
{\bf Proof of Theorem~\ref{m_thm1}} In accordance with Assumption~1, we denote by $M_f$ a common upper bound of the densities $f_{\epsilon_1}$, $f_{\epsilon_2}$, $f_1$ and
$f_2$.

For a given time
transformation $g\in\G_0$, from (\ref{cr1}), we have
\begin{equation}
L_n(g)=\frac{N_n(g)}{D_n(g)}\label{Cn},
\end{equation}
where
\begin{eqnarray}
N_n(g)&=&
\frac{1}{n_1 n_2}
\displaystyle\sum\limits_{i=1}^{n_1}
\displaystyle\sum\limits_{j=1}^{n_2}
\frac{1}{h_1}K_1\left(\frac{{t_i} - g({s_j})}{h_1}\right)
\frac{1}{h_2}K_2\left(\frac{{Y_{t_i}} - Y_{s_j}'}{h_2}\right),\label{Nn}\\
D_n(g)&=&
\frac{1}{n_1 n_2}\displaystyle\sum\limits_{i=1}^{n_1}
\displaystyle\sum\limits_{j=1}^{n_2}
\frac{1}{h_1}K_1\left(\frac{{t_i} - g(s_j)}{h_1}\right)\label{Dn}.
\end{eqnarray}
We first establish that
\begin{eqnarray}
N_n(g)\!\!\!\!\! &\conv{P}&\!\!\!\!\!\intinfinf\intzeroinf f_1(g(y))f_2(y)
f_{\epsilon_1}(v-m(g(y))+m(\g(y)))f_{\epsilon_2}(v) dydv\label{eq_Nn_cov}\\
D_n(g)\!\!\!\!\! &\conv{P}&\!\!\!\!\!\intzeroinf f_1(g(y))f_2(y)dy\label{eq_Dn_cov}.
\end{eqnarray}
The proof is then completed by using the continuous mapping theorem of
convergence in probability
 (see~\citet{Billingsley_1985}).

\noindent
Note that, from (\ref{Nn}), we have
\begin{eqnarray*}
E[N_n(g)]&=&\frac{1}{n_1 n_2}
\displaystyle\sum\limits_{i=1}^{n_1}
\displaystyle\sum\limits_{j=1}^{n_2}
E\left[\frac{1}{h_1}K_1\left(\frac{{t_i} - g({s_j})}{h_1}\right)
\frac{1}{h_2}K_2\left(\frac{{Y_{t_i}} - Y_{s_j}'}{h_2}\right)\right].
\end{eqnarray*}
From the description of the model~(\ref{model1}), we have
\begin{eqnarray*}
E[\Nn{g}]&=&\intinfinf\intinfinf\intzeroinf\intzeroinf\frac{1}{h_1}
K_1\left(\frac{x-
g(y)}{h_1}\right)\\
&\times&\frac{1}{h_2}K_2\left(\frac{m(x)-m(\g(y))+u-v}{h_2}\right)
f_1(x)f_2(y) f_{\epsilon_1}(u)f_{\epsilon_2}(v)dxdydudv.
\end{eqnarray*}

By making the transformations
$z_1=\frac{x-g(y)}{h_1}$ and $z_2=\frac{m(x)- m(\g(y))+u-v}{h_2}$, we have
\begin{eqnarray*}
E[\Nn{g}]&=&
\int_{-\infty}^{\infty}\int_{0}^{\infty}\int_{-\infty}^{\infty}\int_{-\infty}^{
\infty}S_n(z_1,z_2,y,v)
dz_1dz_2dydv
\end{eqnarray*}
where
\begin{eqnarray*}
S_n(z_1,z_2,y,v)&=&I_{(-g(y)/h_1,\infty)}(z_1)K_1\left(z_1\right)
K_2\left(z_2\right)f_1(g(y)+z_1h_1)\notag\\
&&\times f_2(y)
f_{\epsilon_1}(v-m(g(y)+z_1h_1)+m(\g(y))+z_2h_2)f_{\epsilon_2}
(v).
\end{eqnarray*}
As $g$ is a positive and increasing function, any given real $z_1$ is contained
in $(-g(y)/h_1,\infty)$ for sufficiently small $h_1$. By using Assumptions~1,~3
and~4 and the fact that $m$ is a continuous function,
for any fixed $(z_1,z_2,y,v)$, we have
\begin{eqnarray}\label{Sn_limit}
\lim_{n\rightarrow
\infty}S_n(z_1,z_2,y,v)&=&K_1\left(z_1\right)K_2\left(z_2\right)f_1(g(y))f_2(y)\nonumber\\
&\times & f_{\epsilon_1}(v-m(g(y))+m(\g(y)))f_{\epsilon_2}(v).
\end{eqnarray}
Note that, from Assumption 1, we have
\begin{eqnarray}\label{Sn_bd}
0\leq S_n(z_1,z_2,y,v)\leq M_f^2K_1(z_1)K_2(z_2)f_2(y)f_{\epsilon_2}(v).
\end{eqnarray}
Assumption~2 ensures that the bounding function on the right hand side of
(\ref{Sn_bd}) is integrable. Then, by using the Dominated Convergence Theorem (DCT),
we have
\begin{eqnarray}\label{Limit_Nn}
    \lim_{n\rightarrow\infty}E[\Nn{g}]&=&\intinfinf\intzeroinf f_1(g(y))f_2(y)\nonumber\\
    &\times & f_{\epsilon_1}(v-m(g(y))+m(\g(y)))f_{\epsilon_2}(v) dydv.
\end{eqnarray}
From Lemma~\ref{Nn_proof}, proved below, we have $\lim_{n\rightarrow\infty}Var[N_n(g)]=0$. This establishes (\ref{eq_Nn_cov}).

\noindent
We now turn to $D_n(g)$. By using (\ref{Dn}) and making the transformation $z=\frac{
x-g(y)}{h_1}$, we have
\begin{eqnarray*}
E[\Dn{g}]&=&\int_0^\infty\int_{-\infty}^{\infty}R_n(z,y)dzdy,
\end{eqnarray*}
where
\begin{eqnarray}\label{Rn_exp}
 R_n(z,y)=I_{\left(\frac{-g(y)}{h_1},\infty\right)}(z)K_1(z)f_1(g(y)+zh_1)f_2(y).
\end{eqnarray}
From (\ref{Rn_exp}) and Assumptions~1,~3~and~4, we have, for every fixed $z$ and $y$,
\begin{eqnarray*}
 \lim_{n\rightarrow\infty}R_n(z,y)=K_1(z)f_1(g(y))f_2(y).
\end{eqnarray*}
By Assumption~1, we have the dominance
\begin{eqnarray}\label{Rn_bd}
 0\leq R_n(z,y) \leq M_fK_1(z)f_2(y).
\end{eqnarray}
Assumption~1 and~2 ensure that the bounding function on the right hand side of (\ref{Rn_bd}) is integrable. Thus, by applying DCT we have
\begin{eqnarray}\label{Limit_Dn}
 \lim_{n\rightarrow\infty}E[\Dn{g}]=\int_0^\infty f_1(g(y))f_2(y)dy.
\end{eqnarray}

Now from Lemma~\ref{Nn_proof}, proved below, we have $\lim_{n\rightarrow\infty}Var[\Dn{g}]=0$.
This
establishes (\ref{eq_Dn_cov}) and completes the proof. $\hfill \Box$
\bigskip
\begin{lemma}\label{Nn_proof}
Under Assumptions 1, 2, 3 and 4, for any $g\in\G_0$, we have
\begin{eqnarray}
\lim_{n\rightarrow\infty} Var[N_n(g)]&=&0,\label{var_st_1}\\
\lim_{n\rightarrow\infty} Var[D_n(g)]&=&0,\label{var_st_2}
\end{eqnarray}
where $N_n(g)$ and $D_n(g)$ are defined in (\ref{Nn}) and (\ref{Dn}),
respectively.
\end{lemma}
\medskip
\noindent
{\bf Proof of Lemma~\ref{Nn_proof}}
From (\ref{Nn}), we have
\begin{eqnarray}\label{Var_Nn_1step}
&&\hskip-20pt
Var(\Nn{g})\notag\\&=&\frac{1}{(n_1n_2h_1h_2)^2}\sum_{i=1}^{n1}\sum_{j=1}^{n2}\sum_{
i'=1}^{
n1}\sum_{j'=1}^{n2}Cov\bigg \{K_1\left(\frac{t_i- g(s_j)}{h_1}\right)
K_2\left(\frac{Y_{t_i}-Y_{s_j}'}{h_2}\right),\notag\\
&&\qquad
K_1\left(\frac{t_{i'}- g(s_{j'})}{h_1}\right)
K_2\left(\frac{Y_{t_{i'}}-Y_{s_{j'}}'}{h_2}\right)\bigg\}\notag\\
&=& V_1+V_2+V_3+V_4,
\end{eqnarray}
where
\begin{eqnarray}
    V_1&=&\frac{1}{(n_1n_2h_1h_2)^2}\sum_{i=1}^{n1}\sum_{j=1}^{n2}Var\bigg\{
    K_1\left(\frac{t_i- g(s_j)}{h_1}\right)
    K_2\left(\frac{Y_{t_i}-
    Y_{s_j}'}{h_2}\right)\bigg\},\label{eq_V_1} \\
    V_2&=&\frac{1}{(n_1n_2h_1h_2)^2}\sum_{i=1}^{n1}\sum_{j=1}^{n2}\sum^{n1}_{
    \substack{i'=1 \\ (\neq i)}}Cov\bigg\{K_1\left(\frac{t_i- g(s_j)}{h_1}\right)
    K_2\left(\frac{Y_{t_i}-
    Y_{s_j}'}{h_2}\right),\notag\\
    &&\qquad K_1\left(\frac{t_{i'}-g(s_{j})}{h_1}\right)
    K_2\left(\frac{Y_{t_{i'}}-Y_{s_j}'}{h_2}\right)\bigg\},\label{eq_V_2}
\end{eqnarray}
\begin{eqnarray}
    V_3&=&\frac{1}{(n_1n_2h_1h_2)^2}\sum_{i=1}^{n1}\sum_{j=1}^{n2}\sum^{n2}_{\substack{
    j'=1\\ (\neq j)}} Cov\bigg\{K_1\left(\frac{t_i-g(s_j)}{h_1}\right)
    K_2\left(\frac{Y_{t_i}-
    Y_{s_j}'}{h_2}\right),\notag\\
    &&\qquad K_1\left(\frac{t_{i}- g(s_{j'})}{h_1}\right)
    K_2\left(\frac{Y_{t_i}-Y_{s_{j'}}'}{h_2}\right)\bigg\},\notag\\
    V_4&=&\frac{1}{(n_1n_2h_1h_2)^2}\sum_{i=1}^{n1}\sum_{j=1}^{n2}\sum^{n1}_{\substack{
    i'=1 \\ (\neq i)}}\sum^{n2}_{\substack{j'=1\\ (\neq
    j)}} Cov\bigg\{K_1\left(\frac{t_i- g(s_j)}{h_1}\right)
    K_2\left(\frac{Y_{t_i}-Y_{s_j}'}{h_2}\right)\notag\\
    &&\qquad K_1\left(\frac{t_{i'}- g(s_{j'})}{h_1}\right)
    K_2\left(\frac{Y_{t_{i'}}-Y_{s_{j'}}'}{h_2}\right)\bigg\}.
\end{eqnarray}
We consider the convergence of each term on the right hand side of
(\ref{Var_Nn_1step}) separately.
By using (\ref{model1}) and (\ref{eq_V_1}), we have
\begin{eqnarray*}
V_1&=&\frac{1}{n_1n_2h_1h_2}V_{11}-\frac1{n_1n_2}E^2[N_n(g)],
\end{eqnarray*}
where
\begin{eqnarray*}
V_{11}&=&\int_{-\infty}^{
\infty}\int_{-\infty}^{\infty}\int_{0}^{\infty}\int_{0}^{\infty}\frac1{h_1}
K_1^2\left(\frac{x-g(y)}{h_1}\right)
\frac1{h_2}K_2^2\left(\frac{m(x)-
m(g_0(y))+u-v}{h_2}\right)\nonumber\\&&\hskip1.4in\times
f_1(x)f_2(y) f_{\epsilon_1}(u)f_{\epsilon_2}(v)dxdydudv\label{exp_V11}.
\end{eqnarray*}
By making the transformations $z_1=\frac{x-g(y)}{h_1}$
and $z_2=\frac{m(x)- m(\g(y))+u-v}{h_2}$, we have
\begin{eqnarray}
V_{11}=
\int_{-\infty}^{\infty}\int_{0}^{\infty}\int_{-\infty}^{\infty}\int_{-\infty}^{
\infty }
I_{(\frac{-g(y)}{h_1},\infty)}(z_1)K_1^2(z_1)K_2^2(z_2)
f_1(g(y)+z_1h_1)\nonumber\\
\hskip 0.25in\times f_2(y)
f_{\epsilon_1}(v-m(g(y)+z_1h_1)+m(\g(y))+z_2h_2)f_{\epsilon_2}(v)
dz_1dz_2dydv.\label{eq_v11}
\end{eqnarray}
From Assumptions~1,~3,~and~4 and the continuity of $m$, for any fixed real
$(z_1,z_2,y,v)$, a similar argument as given for (\ref{Sn_limit})
shows that
the integrand function on the right hand side of~(\ref{eq_v11}) converges, as $n\rightarrow\infty$, to
\begin{eqnarray*}
K_1^2(z_1)K_2^2(z_2)f_1(g(y))f_2(y)
f_{\epsilon_1}(v-m(g(y))+m(\g(y)))f_{\epsilon_2}(v).
\end{eqnarray*}

We have the dominance of the integrand function on the right hand side of
(\ref{eq_v11}) by the integrable function
\begin{eqnarray*}
M_f^2K_1^2(z_1)K_2^2(z_2)f_2(y)f_{\epsilon_2}(v).
\end{eqnarray*}
By using DCT and convergence of the integrand on the right hand
side of (\ref{eq_v11}), we have
\begin{eqnarray*}
\lim_{n\rightarrow\infty}V_{11}&=&\int_{-\infty}^{\infty}K_1^2(z_1)dz_1\int_{
-\infty}^{\infty} K_2^2(z_2)dz_2\\
&&\qquad\times \int_ { -\infty}^{\infty}\int_{0}^{\infty}
f_1(g(y))f_2(y)f_{\epsilon_1}(v-m(g(y))+m(\g(y)))f_{\epsilon_2}(v) dydv.
\end{eqnarray*}
Now, from (\ref{Limit_Nn}), we have
$$\frac1{n_1n_2}E^2(N_n(g))=O\left(\frac{1}{n_1n_2}\right).$$
Thus, we have
\begin{equation}\label{V1_final}
V_1=O\left(\frac1{n_1n_2h_1h_2}\right)+O\left(\frac1{n_1n_2}
\right)=O\left(\frac1{n_1n_2h_1h_2}\right).
\end{equation}
We now consider the term $V_2$. From (\ref{eq_V_2}) and (\ref{model1}), we
have
\begin{eqnarray}\label{V2_exp}
V_2&=&\frac{n_1-1}{n_1n_2} V_{21}- \frac{n_1-1}{n_1n_2} E^2(N_n(g)),
\end{eqnarray}
Where
\begin{eqnarray}\label{eq_v21}
    V_{21}&=&\int_{-\infty}^{\infty}\int_{
    -\infty}^{\infty}\int_{-\infty}^{\infty}\int_{0}^{\infty}\int_{0}^{\infty}\int_{
    0}^{\infty}
    \frac1{h_1}K_1\left(\frac{x-g(y)}{h_1}\right)\notag\\
    &\times &\frac1{h_2}K_2\left(\frac{m(x)- m(\g(y))+u-v}{h_2}\right)\notag\\
    &\times &\frac1{h_1}K_1\left(\frac{x'-g(y)}{h_1}\right)
    \frac1{h_2}K_2\left(\frac{m(x')-
    m(\g(y))+u'-v}{h_2}\right)\notag\\
    &\times &f_1(x)f_1(x')f_2(y)f_{\epsilon_1}(u)f_{\epsilon_1}(u')f_{
    \epsilon_2}(v)dxdx'dydudu'dv.
\end{eqnarray}
By making the transformations $z_1=\frac{x-g(y)}{h_1}$, $z_2=\frac{x'-g(y)}{h_1}$,
$z_3=\frac{m(x)- m(\g(y))+u-v}{h_2}$, and $z_4=\frac{m(x')-
m(\g(y))+u'-v}{h_2}$, the integrand on the right hand side of (\ref{eq_v21}) is
\begin{eqnarray}\label{integrand_v21}
&&I_{(-g(y)/h_1,\infty)}(z_1)K_1(z_1)
I_{(-g(y)/h_1,\infty)}(z_2)K_1(z_2)
K_2(z_3)K_2(z_4)f_1(g(y)+z_1h_1)\notag\\
&&\times\;f_1(g(y)+z_2h_1)
f_2(y) f_{\epsilon_1}(v-m(g(y)+z_1h_1)+m(g_0(y))+h_2z_3)\notag\\
&&\times\;f_{\epsilon_1}(v-m(g(y)+z_2h_1)+m(g_0(y))+h_2z_4)
f_{\epsilon_2}(v).
\end{eqnarray}
From Assumptions~1,~3,~and~4 and the continuity of $m$ it follows via a
similar argument for (\ref{Sn_limit}) that for any
$z_1,z_2,z_3,z_4,y$ and $v$ the above function converges, as
$n\rightarrow\infty$, to
\begin{eqnarray*}
K_1(z_1)K_1(z_2)K_2(z_3)K_2(z_4) f_1^2(g(y)) f_2(y)
f_{\epsilon_1}^2(v-m(g(y))+m(g_0(y))) f_{\epsilon_2}(v).
\end{eqnarray*}
The integrand function on the right hand side of (\ref{integrand_v21})
is dominated by the integrable function
\begin{eqnarray*}
M_f^4K_1(z_1)K_1(z_2)K_2(z_3)K_2(z_4)f_2(y)f_{\epsilon_2}(v).
\end{eqnarray*}
By using Assumption~2 and the convergence of (\ref{integrand_v21}), and
applying DCT on the right hand side of (\ref{eq_v21}), we have
\begin{eqnarray*}
\lim_{n\rightarrow\infty}V_{21}&=&\int_{0}^\infty\int_{-\infty}^\infty
f_1^2(g(y))f_2(y)
f_{\epsilon_1}^2(v-m(g(y))+m(g_0(y)))f_{\epsilon_2}(v)dvdy.
\end{eqnarray*}
Now, from (\ref{Limit_Nn}), the second term on the right hand side of (\ref{V2_exp}) turns out to be
$$\frac{n_1-1}{n_1n_2}E^2(N_n(g))=O\left(\frac{1}{n_2}\right).$$
It follows that
\begin{equation*}
 n_2V_2\rightarrow\int_{0}^\infty\int_{-\infty}^\infty f_1^2(g(y))f_2(y)
f_{\epsilon_1}^2(v-m(g(y))+m(g_0(y)))f_{\epsilon_2}(v)dvdy - E^2(N_n(g))
\end{equation*}
i.e.,
\begin{equation}\label{V2_final}
V_2=O\left(\frac1{n_2}\right).
\end{equation}
By using a similar argument as for the term $V_2$, we have
\begin{equation*}
 n_1V_3\rightarrow\int_{0}^\infty\int_{-\infty}^\infty
\frac{f_2^2(y)}{g'(y)}f_1(g(y))
f_{\epsilon_1}(v-m(g(y))+m(g_0(y)))f_{\epsilon_2}^2(v) dvdy - E^2(N_n(g))
\end{equation*}
i.e.,

\begin{equation}\label{V3_final}
V_3=O\left(\frac1{n_1}\right).
\end{equation}
Finally, we consider the term $V_4$. By using the model specification (\ref{model1}), we have
\begin{equation}\label{V4_final}
V_4=0.
\end{equation}
The proof of (\ref{var_st_1}) is completed from (\ref{V1_final}), (\ref{V2_final}), (\ref{V3_final}), (\ref{V4_final}) and by using Assumptions~3 and~4.

\noindent
We now compute $Var[D_n(g)]$. Note that, from (\ref{Dn}), we have
\begin{eqnarray*}
Var[\Dn{g}]&=&T_1+T_2+T_3+T_4,\quad
\end{eqnarray*}
Where
\begin{eqnarray*}
T_1&=&\frac{1}{(n_1n_2h_1)^2}\sum_{i=1}^{n1}\sum_{j=1}^{n2}
Var\left\{K_1\left(\frac {
t_i-g(s_j)}{h_1}\right)\right\},\\
T_2&=&\frac{1}{(n_1n_2h_1)^2}\sum_{i=1}^{n1}\sum_{j=1}^{n2}\sum^{n1}_{
\substack {
i'=1\\ (\neq
i)}}Cov\left\{K_1\left(\frac{t_i-g(s_j)}{h_1}\right),K_1\left(\frac{t_{i'}
-g(s_j) } {
h_1}\right)\right\},\\
T_3&=&\frac{1}{(n_1n_2h_1)^2}\sum_{i=1}^{n1}\sum_{j=1}^{n2}\sum^{n2}_{
\substack {
j'=1\\ (\neq
j)}}Cov\left\{K_1\left(\frac{t_i-g(s_j)}{h_1}\right),K_1\left(\frac{t_i-g(s_{j'}
) } {
h_1}\right)\right\},\\
T_4&=&\frac{1}{(n_1n_2h_1)^2}\sum_{i=1}^{n1}\sum_{j=1}^{n2}\sum_{\substack{
i'=1\\
(\neq i)}}^{n1}\sum_{\substack{j'=1\\(\neq j)}}^{n2}
Cov\left\{K_1\left(\frac{t_i-
g(s_j)}{h_1}\right),K_1\left(\frac{t_{i'}-g(s_{j'})}{h_1}\right)\right\}.
\end{eqnarray*}
We consider the convergence of the terms $T_1, T_2$, $T_3$ and $T_4$ separately.
Consider the term $T_1$. By making the transformation
$z=\frac{x-g(y)}{h_1}$ and by using the model specification, we have
\begin{eqnarray}\label{T1_exp}
T_1&=&\frac{1}{n_1n_2h_1}T_{11}-\frac{1
}{n_1n_2}E^2[D_n(g)],
\end{eqnarray}
where
\begin{eqnarray}\label{T_11}
T_{11}=\int_{0}^\infty\int_{-\infty}^\infty
I_{(-g(y)/h_1,\infty)}(z)K_1^2(z)f_1(g(y)+h_1z)f_2(y)dzdy.
\end{eqnarray}
Note that the integrand on the right hand side of (\ref{T_11}) is bounded by the
integrable function $M_fK_1^2(z)f_2(y)$. Further, a similar argument, as used
for the $Var(N_n)$, shows that, for any given $y$ and $z$, the integrand
function converges to $K_1^2(z)f_1(g(y))f_2(y)$ as $n\rightarrow\infty$. Thus,
by applying DCT and Assumptions~3 and~4, we have
\begin{eqnarray*}
\lim_{n\rightarrow\infty}T_{11}=\int_{-\infty}^\infty K_1^2(z) dz
\int_{0}^\infty f_1(g(y))f_2(y)dy.
\end{eqnarray*}
Now, from (\ref{Limit_Dn}), the second term on the right hand side of (\ref{T1_exp}) turns out to be
$$\frac{1}{n_1n_2}E^2[D_n(g)]=O\left(\frac{1}{n_1n_2}\right).$$
Thus, we have
$$T_1=O\left(\frac{1}{n_1n_2h_1}\right)+O\left(\frac{1}{n_1n_2}
\right)=O\left(\frac{1}{n_1n_2h_1}\right).$$

\noindent
We now consider the term $T_2$.
By making the transformations $z=\frac{x-g(y)}{h_1}$ and $z'=\frac{x'-g(y)}{h_1}$, we
have from (\ref{model1})
\begin{eqnarray}\label{exp_T2}
T_2&=&\frac{n_1-1}{n_1n_2}T_{21}-\frac{n_1-1}{n_1n_2}E^2[D_n(g)],
\end{eqnarray}
where
\begin{eqnarray}\label{T21}
T_{21}&=&\int_0^\infty\int_{-\infty}^\infty\int_{-\infty}
^\infty
I_{(-g(y)/h_1,\infty)}(z)K_1(z)I_{(-g(y)/h_1,
\infty) }(z')K_1(z') \notag\\
&&\times f_1(g(y)+zh_1)f_1(g(y)+z'h_1)f_2(y)dzdz'dy.
\end{eqnarray}
Note that, the integrand on the right hand side of (\ref{T21}) is bounded by the
integrable function $M_f^2K_1(z)K_1(z')f_2(y)$. A similar argument as used for
the convergence of the term $T_{11}$ shows that, for any given $z,z'$ and $y$,
the integrand function converges to $
K_1(z)K_1(z')f_1^2(g(y))f_2(y)$, as $n\rightarrow\infty$. Thus, by applying
DCT and Assumption~~2,~3
and~4, we have
\begin{eqnarray*}
\lim_{n\rightarrow\infty}T_{21}=
\int_{0}^\infty f_1^2(g(y)) f_2(y)dy.
\end{eqnarray*}
Now, from (\ref{Limit_Dn}), the second term on the right hand side of (\ref{exp_T2}) turns out to be
$$\frac{n_1-1}{n_1n_2}E^2(D_n(g))=O\left(\frac{1}{n_2}\right).$$
Thus, we have
$$
n_2T_2\rightarrow\int_{0}^\infty f_1^2(g(y)) f_2(y)dy-E^2(D_n(g))
$$
i.e.,
$$T_2=O\left(\frac{1}{n_2}\right).$$

We now consider the term $T_3$. A similar argument, as used for the convergence of the term $T_2$, shows that
$$
n_1T_3\rightarrow\int_{0}^\infty
\frac{f_1(g(y))f_2^2(y)}{g'(y)}dy-E^2(D_n(g))
$$
i.e.,
$$T_3=O\left(\frac{1}{n_1}\right).$$

\noindent
Finally, by using the model specification, we have $T_4=0$. This completes the proof of (\ref{var_st_2}).$\hfill \Box$

\bigskip
\noindent
{\bf Proof of Theorem~\ref{m_thm1.5}}
Let us denote the convolution of the densities $f_{\epsilon_1}$ and $f_{\epsilon_2}$ by $f_{\epsilon_1+\epsilon_2}$. We first show that $f_{\epsilon_1+\epsilon_2}$ is strictly unimodal at zero. Indeed, for $u>0$, we observe from Assumption~1 that
\begin{eqnarray*}
f_{\epsilon_1+\epsilon_2}(u)
&=&\intinfinf f_{\epsilon_1}(v)f_{\epsilon_2}(u-v)dv\\
&=&\intzeroinf f_{\epsilon_1}(v)f_{\epsilon_2}(u-v)dv+\int_{0}^{\infty} f_{\epsilon_1}(v)f_{\epsilon_2}(u+v)dv.\\
\end{eqnarray*}

Because of the strict unimodality of $f_{\epsilon_2}$, for $u_2>u_1>0$, we have
\begin{eqnarray*}
&&\hskip-22pt f_{\epsilon_1+\epsilon_2}(u_2) - f_{\epsilon_1+\epsilon_2}(u_1)\\
&=&\int_0^{\infty} f_{\epsilon_1}(v) \left[\left\{f_{\epsilon_2}(u_2-v) - f_{\epsilon_2}(u_1-v)\right\}
+ \left\{f_{\epsilon_2}(u_2+v)-f_{\epsilon_2}(u_1+v)\right\}\right]dv\\
&<&0.
\end{eqnarray*}

By a similar argument, the same inequality holds for $u_2< u_1 <0$. Thus, $f_{\epsilon_1+\epsilon_2}$ is strictly unimodal at zero.

Now observe from \eqref{exp_C} and Assumption~1 that
\begin{eqnarray*}
L(g)&=&\frac{\intinfinf\intzeroinf f_1(g(y))f_2(y)f_{\epsilon_1}(v-m(g(y))+m(g_0(y)))f_{\epsilon_2}(v) dydv}{\intzeroinf f_1(g(y))f_2(y)dy}\\
&=&\frac{\intzeroinf f_1(g(y))f_2(y)f_{\epsilon_1+\epsilon_2}(m(g(y))-m(g_0(y))) dy}{\intzeroinf f_1(g(y))f_2(y)dy}.
\end{eqnarray*}
In particular, $L(g_0)=f_{\epsilon_1+\epsilon_2}(0)$. Thus,
\begin{eqnarray*}
L(g_0)-L(g)
&=&f_{\epsilon_1+\epsilon_2}(0)
-\frac{\intzeroinf f_1(g(x))f_2(x)f_{\epsilon_1+\epsilon_2}(m(g(x))-m(g_0(x))) dx}{\intzeroinf f_1(g(x))f_2(x)dy}\\
&=&\frac{\intzeroinf f_1(g(x))f_2(x)\left[f_{\epsilon_1+\epsilon_2}(0)-f_{\epsilon_1+\epsilon_2}(m(g(x))-m(g_0(x)))\right] dx}{\intzeroinf f_1(g(x))f_2(x)dy}.
\end{eqnarray*}
Unimodality of $f_{\epsilon_1+\epsilon_2}$ at~0 implies that
$$[f_{\epsilon_1+\epsilon_2}(0)-f_{\epsilon_1+\epsilon_2}(m(g(x))-m(g_0(x)))]\ge0\quad\mbox{for all }x.$$
This inequality proves part~(a).

In order that the last expression for $L(g_0)-L(g)$ happens to be zero for some $g\in\G_0$, the above difference must be equal to zero for all $x$ such that $f_1(g(x))f_2(x)>0$, i.e., for $x\in S_g\cap S_{g_0}$, where $S_g$ is as defined at the beginning of Section~3. Since $f_{\epsilon_1+\epsilon_2}$ is strictly unimodal at 0, this requirement reduces to $$m(g(x))=m(g_0(x))\qquad \forall x\in S_g\cap S_{g_0}.$$
It follows from Theorem~\ref{m_thm0} that $S_g=S_{g_0}$ and $g(x)= g_0(x)$ $\forall x\in S_{g_0}$. This completes the proof of part~(b).$\hfill \Box$

\bigskip
\noindent
{\bf Proof of Theorem~\ref{m_thm2}}
In accordance with Assumptions~2 and ~2A, let the positive real numbers $c,M_K,M'_K$ be such
that $0<c\leq K_i(x)\leq M_K$ and $|K_{i}'(x)|\leq M_{K}'$ for $i=1,2$.

We first obtain a stochastic upper bound on the variation in $L_n(\cdot)$. From
(\ref{Cn}), for any given $\tilde{g},g\in\G$, we have
\begin{eqnarray*}
|L_n(\tilde{g})-L_n(g)|
&=&\left|\frac{\Nn{\tilde{g}}}{\Dn{\tilde{g}}}-\frac{\Nn{g}}{\Dn{g}}
\right|\\
&=&\frac{|\Nn{\tilde{g}}\Dn{g}-\Nn{g}\Dn{\tilde{g}}|}{\Dn{\tilde{g}}\Dn{g}
}.
\end{eqnarray*}
From (\ref{Dn}) and Assumption~2A, we have
$\Dn{g}\geq\frac{c}{h_1}$. Therefore,
\begin{eqnarray}\label{Csup_step1}
    &&\hskip-20pt|L_n(\tilde{g})-L_n(g)|\notag\\
    &\leq &\frac{h_1^2}{c^2}\left|\Nn{\tilde{g}}\Dn{g}-\Nn{g}\Dn{\tilde{g}}
    \right|\notag\\
    &\leq &\frac{h_1^2}{c^2}\left\{\Dn{\tilde{g}}\left|\Nn{\tilde{g}}-\Nn{g}
    \right|+\Nn{\tilde{g}}\left|\Dn{\tilde{g}}-\Dn{g}\right|\right\}.
\end{eqnarray}

We now compute the upper bounds for both the terms on the right hand side of
(\ref{Csup_step1}).
Note that, from (\ref{Nn}), we have
\begin{eqnarray*}
    &&\hskip-10pt|\Nn{\tilde{g}}-\Nn{g}|\\
    &=&\left|\frac{1}{n_1n_2}\hskip-2pt\sum_{i=1}^{n1}\sum_{j=1
    }^{n2}\frac{1}{h_1}\left\{
    K_{1}\left(\frac{t_i-\tilde{g}(s_j)}{h_1}\right)-K_{1}
    \left(\frac{
    t_i-g(s_j) }{h_1}\right)\right\}\frac{1}{h_2}K_{2}\left(\frac{
    Y_{t_i}-Y'_{s_j}}{h_2}\right)\right|\\
    &\leq&\frac{1}{n_1n_2h_1h_2}\sum_{i=1}^{n1}\sum_{j=1
    }^{n2}K_{2}\left(\frac{Y_{t_i}-Y'_{s_j}}{h_2}\right)\left|
    K_{1}\left(\frac{t_i-\tilde{g}(s_j)}{h_1}\right)-K_{1}\left(\frac{
    t_i-g(s_j)}{h_1}\right)\right|.
\end{eqnarray*}
By using the mean value theorem, we have
\begin{eqnarray}\label{Nn_sup_step1}
    &&\hskip-20pt|\Nn{\tilde{g}}-\Nn{g}|\notag\\
    &\le & \frac{1}{n_1n_2h_1h_2}\sum_{i=1}^{n1}\sum_{j=1
    }^{n2}K_{2}\!\left(\frac{Y_{t_i}-Y'_{s_j}}{h_2}\right)\notag\\
    &\times & \left|K_{1}^{'}(x_0(t_i,s_j,\tilde{g},g))\right|\left|\frac{g
    (s_j)-\tilde{g}(s_j)}{h_1}
    \right|,
\end{eqnarray}
where $x_0(t_i,s_j,\tilde{g},g)\in \left(\min\left(\frac{t_i-\tilde{g}(s_j)}{h_1},\frac{
t_i-g(s_j)}{h_1}\right),\max\left(\frac{t_i-\tilde{g}(s_j)}{h_1},\frac{
t_i-g(s_j)}{h_1}\right)\right)$.
Now, from (\ref{Nn_sup_step1}) and Assumption~2A, we
have
\begin{eqnarray}\label{Nn_sup_step2}
|\Nn{\tilde{g}}-\Nn{g}|
&\leq&\frac{M_K^{'}}{h_1^2}\cdot
\|g-\tilde{g}\|\cdot U_n,\label{uceq1}
\end{eqnarray}
where
\begin{equation}\label{Un}
U_n=\frac{1}{n_1n_2h_2}\sum_{i=1}^{n1}
\sum_{j=1}^{n2}K_{2}\left(\frac{Y_{t_i}-Y'_{s_j}}{h_2}\right).
\end{equation}
We now turn to the second term on the right hand side of (\ref{Csup_step1}).
From (\ref{Dn}), we have
\begin{eqnarray*}
|\Dn{\tilde{g}}-\Dn{g}|
&\leq&\frac{1}{n_1n_2h_1}\sum_{i=1}^{n1}\sum_{j=1
}^{n2}\left|K_{1}\left(\frac{t_i-\tilde{g}(s_j)}{h_1}\right)-K_{1}\left(\frac{
t_i-g(s_j) }{h_1}\right)\right|.
\end{eqnarray*}
From  Assumption~2A and the mean value theorem,
we have
\begin{eqnarray}
|\Dn{\tilde{g}}-\Dn{g}|&\le&\frac{1}{n_1n_2h_1}\sum_{i=1}^{n1}\sum_{j=1
}^{n2}\left|K_{1}^{'}(x_1(t_i,s_j,\tilde{g},g))\right|\left|\frac{g
(s_j)-\tilde{g}(s_j)}{h_1
}\right|\nonumber\\
&\leq&\frac{M_K^{'}}{h_1^2}\cdot \|g-\tilde{g}\|,
\label{Dn_sup_step1}
\end{eqnarray}
where where $x_1(t_i,s_j,\tilde{g},g)\in \left(\min\left(\frac{t_i-\tilde{g}(s_j)}{h_1},\frac{
t_i-g(s_j)}{h_1}\right),\max\left(\frac{t_i-\tilde{g}(s_j)}{h_1},\frac{
t_i-g(s_j)}{h_1}\right)\right)$.
\noindent
Now, by using (\ref{Nn_sup_step2}), (\ref{Dn_sup_step1}) and
(\ref{Csup_step1}), we have
\begin{eqnarray}\label{Csup_step2}
|L_n(\tilde{g})-L_n(g)|
&\le&B_n(\tilde{g})\cdot\|g-\tilde{g}\|,
\end{eqnarray}
where $$B_n(\tilde{g})=\frac{M_K^{'}}{c^2}\left\{\Nn{\tilde{g}}+
U_n \cdot \Dn{\tilde{g}}\right\}.$$
The expression on the right hand side of (\ref{Csup_step2}) gives an upper bound on the change of the functional $L_n(\cdot)$ with change in time transformation functions in $\G$.

\noindent
Note that
\begin{eqnarray}\label{cn_sup_ie_1}
|L_n(g)-L(g)|
&\leq&|L_n(g)-L_n(\tilde{g})|
+
|L(\tilde{g})-L(g)|+|L_n(\tilde{g})-L(\tilde{g})|,
\end{eqnarray}
where $L(.)$ is defined as in (\ref{exp_C}).


\noindent
Set $\epsilon>0$.
\noindent
Lemma \ref{lem_UC}, proved below, implies that there
exists $\delta_{\epsilon}>0$ such
that
\begin{equation}{\label{m_thm2_5}}
 \|g-\tilde{g}\|<\delta_{\epsilon}\;\mbox{implies}\;|L(g)-L(\tilde{g}
)|<\frac{\epsilon}{3}.
\end{equation}

\noindent
Theorem~\ref{m_thm1} and Lemma \ref{lem_Un}, proved below, implies that for all $\tilde{g}$
there
exists $M_{\tilde{g}}$ such that $B_n(\tilde{g})\conv{P}M_{\tilde{g}}$, which
ensures
\begin{equation}{\label{m_thm2_6}}
P\left(B_n(\tilde{g}
)>\max\left\{\frac{\epsilon}{3\delta_{\epsilon}},2M_{\tilde{g}}\right\}\right)
\rightarrow0.
\end{equation}

\noindent
Define $\mathcal{N}_{\eta}(\tilde{g})=\left\{g:\|g-\tilde{g}\|<\eta\right\}.$
For given $\tilde{g}$, let
\begin{eqnarray}\label{m_thm2_7}
\delta(\tilde{g},\epsilon)=
\left\{
\begin{array}{lll}
\min\left\{\frac{\epsilon}{6
M_{\tilde{g}}},\delta_{\epsilon}\right\}&\mbox{if}& M_{\tilde{g}}>0\\
\delta_{\epsilon}&\mbox{if}& M_{\tilde{g}}=0.
\end{array}\right.
\end{eqnarray}
For $g$ in $\mathcal{N}_{\delta(\tilde{g},\epsilon)}(\tilde{g})$, we have from
(\ref{Csup_step2})
\begin{equation}\label{m_thm2_8}
|L_n(\tilde{g})-L_n(g)|
<\delta(\tilde{g},\epsilon)\cdot
B_n(\tilde{g}).
\end{equation}
Note that $\left\{\mathcal{N}_{\delta(\tilde{g},\epsilon)}(\tilde{g}):\tilde{g}
\in\G\right
\}$ is an open cover of $\G$. By Assumption~5, there exists a finite
sub-cover say
$\left\{\mathcal{N}_{\delta(\tilde{g}_j,\epsilon)}(\tilde{g}_j)\right\}_{j=1
\ldots k_\epsilon}$, with
$\G\subset\cup_{j=1}^{k_{\epsilon}}\mathcal{N}_{\delta(\tilde{g}_j,\epsilon)}
(\tilde { g }_j)$ for some finite $k_\epsilon$.
From (\ref{cn_sup_ie_1}), (\ref{m_thm2_5}), and (\ref{m_thm2_8}) we have,

\begin{eqnarray}\label{m_thm2_9}
&&\hskip-30pt  \sup_{g\in\G}|L_n(g)-L(g)|\nonumber\\
&\leq&\max_{j=1,\ldots,k_\epsilon}\sup_{
g\in\mathcal{N}_{\delta(\tilde{g}_j,\epsilon)}(\tilde{g}_j)}
|L_n(g)-L(g)|\notag\\
&\leq&\max_{j=1,\ldots,k_\epsilon}\left\{
\sup_{
g\in\mathcal{N}_{\delta(\tilde{g}_j,\epsilon)}(\tilde{g}_j)}
|L_n(g)-L_n(\tilde{g}_j)|+\sup_{
g\in\mathcal{N}_{\delta(\tilde{g}_j,\epsilon)}(\tilde{g}_j)}
|L(\tilde{g} _j)-L(g)|\right.\notag\\&&+\left.\sup_{
g\in\mathcal{N}_{\delta(\tilde{g}_j,\epsilon)}(\tilde{g}_j)}
|L_n(\tilde{g}_j)-L(\tilde { g } _j)|\right\}\notag\\
&\leq&\max_{j=1,\ldots,k_\epsilon}\left\{\delta(\tilde{g}_j,
\epsilon)B_n(\tilde{g}_j)+\frac{\epsilon}{3}+|L_n(\tilde{g}_j)-L(\tilde{g}
_j)|\right\}\notag\\
&\leq&\max_{j=1,\ldots,k_\epsilon}\delta(\tilde{g}_j,
\epsilon)B_n(\tilde{g}_j)+\frac{\epsilon}{3}+\sum_{j=1}^{k_\epsilon}|L_n(\tilde{
g } _j)-L(\tilde{g}_j)|.
\end{eqnarray}

\noindent
From (\ref{m_thm2_7}) and (\ref{m_thm2_9}) we have,
\begin{eqnarray}
 &&\hskip-30pt  P\left\{\sup_{g\in\G}
|L_n(g)-L(g)|>\epsilon\right\}\notag\\
&\leq&
P\left\{\max_{j=1,\ldots,k_{\epsilon}}\delta(\tilde { g }
_j,\epsilon)B_n(\tilde{g}_j)>\frac{\epsilon}{3}\right\}+P\left\{\sum_{j=1}^{
k_\epsilon}|L_n(\tilde{
g }_j)-L(\tilde{g}_j)|>\frac{\epsilon}{3}\right\}\notag\\
&\leq&\sum_{j=1}^{k_{\epsilon}}
P\left\{B_n(\tilde{g}_j)>\frac{\epsilon}{3\delta(\tilde { g }
_j,\epsilon)}\right\}+P\left\{\sum_{j=1}^{
k_\epsilon}|L_n(\tilde{
g }_j)-L(\tilde{g}_j)|>\frac{\epsilon}{3}\right\}.\label{m_thm2_10}
\end{eqnarray}

\noindent
Each summand of the first term on the right hand side of
(\ref{m_thm2_10}) goes to zero by (\ref{m_thm2_6}), while the second term goes
to zero by Theorem~\ref{m_thm1}. This completes the proof.
$\hfill\Box$

\bigskip
\begin{lemma}\label{lem_UC}
Under Assumptions~1 and~5 the functional $L(\cdot)$ in (\ref{exp_C}) is
uniformly continuous on $\G$.
\end{lemma}
\medskip
\noindent
{\bf Proof:} Let
\begin{eqnarray}
    N(g)\!\!\!\!\!&=&\!\!\!\!\!\intinfinf\intzeroinf f_1(g(y))f_2(y)
    f_{\epsilon_1}(v-m(g(y))+m(\g(y)))f_{\epsilon_2}(v) dydv,\label{Ng}\\
    D(g)\!\!\!\!\!&=&\!\!\!\!\!\intzeroinf
    f_1(g(y))f_2(y)dy.\label{Dg}
\end{eqnarray}
Then, $L(g)=\frac{N(g)}{D(g)}$.

\noindent
Let $g\in\G$ and $\{g_k\in\G; k=1,2,\ldots\}$ be such that
$\lim_{k\rightarrow\infty}\sup|g_k-g|=0$. Now, from~(\ref{Ng})
\begin{eqnarray}\label{N_1step}
    \lim_{k\rightarrow\infty}N(g_k)&=&\lim_{k\rightarrow\infty}
    \intinfinf\intzeroinf f_1(g_k(y))f_2(y)\notag\\
    &\times &f_{\epsilon_1}(v-m(g_k(y))+m(g_0(y)))f_{\epsilon_2}(v) dydv.
\end{eqnarray}
Note that the integrand on the right hand side of (\ref{N_1step}) is bounded by the integrable function
$M_f^2f_2(y)f_{\epsilon_2}(v).$
Thus, applying DCT, we have
\begin{eqnarray}\label{N_2step}
    \lim_{k\rightarrow\infty}
    N(g_k)&=&\intinfinf\!\intzeroinf
    \left\{\lim_{k\rightarrow\infty} f_1(g_k(y))\right\}
    f_2(y)\notag\\
    &\times&\left\{\lim_{k\rightarrow\infty} f_{\epsilon_1}(v-m(g_k(y))+m(g_0(y)))\right\} f_{\epsilon_2}(v)
    dydv.
\end{eqnarray}
Note that $g_k\rightarrow g$ as $k\rightarrow\infty$ pointwise. By using Assumption~1 and the fact that $m$ is continuous, we have
\begin{eqnarray*}
\lim_{k\rightarrow\infty}f_1(g_k(y))&=& f_1(g(y)),\\
\lim_{k\rightarrow\infty}f_{\epsilon_1}(v-m(g_k(y))+m(g_0(y)))&=&f_{\epsilon_1}(v-m(g(y))+m(g_0(y))).
\end{eqnarray*}
Thus, from (\ref{N_2step}), we have
\begin{eqnarray*}
\lim_{k\rightarrow\infty}N(g_k)&\!\!\!=\!\!\!&\intinfinf\!\intzeroinf f_1(g(y))f_2(y)
f_{\epsilon_1}(v-m(g(y))+m(g_0(y)))f_{\epsilon_2}(v) dydv\notag\\
&\!\!\!=\!\!\!&N(g).
\end{eqnarray*}
This shows that the functional $N(\cdot)$ is continuous on $\G$.

A similar argument shows that
$D(\cdot)$ is also continuous. Further, note from
Assumption~1 and (\ref{Dg}) that $D(g)>0$ for any $g\in \G$. This establishes
that
$L$ is continuous on $\G$. From Assumption~5, $L$ is uniformly continuous on
$\G$. $\hfill \Box$

\bigskip
\begin{lemma}\label{lem_Un}
Let
\begin{equation}\label{Un_exp}
U_n=\frac{1}{n_1n_2h_2}\sum_{i=1}^{n1}
\sum_{j=1}^{n2}K_2\left(\frac{Y_{t_i}-Y'_{s_j}}{h_2}\right).
\end{equation}
Then, under Assumptions~1, 2, 3 and 4,
\begin{eqnarray*}
U_n\conv{P}
\intinfinf\intzeroinf\intzeroinf f_1(x)f_2(y)f_{\epsilon_1}(v-m(x)+m(g_0(y)))f_{
\epsilon_2}(v)dxdydv.
\end{eqnarray*}
\end{lemma}
\medskip
\noindent
{\bf Proof:} From~(\ref{Un_exp}) and~(\ref{model1}), we have
\begin{eqnarray*}
    E(U_n)&=&\intinfinf\intinfinf\intzeroinf\intzeroinf \frac{1}{h_2}K_2\left(\frac{
    m(x)-m(\g(y))+u-v}{h_2}\right)\\
    &\times & f_1(x)f_2(y)f_{\epsilon_1}(u)f_{\epsilon_2}(v) dxdydudv.
\end{eqnarray*}
By making the transformation $w=\frac{m(x)-m(\g(y))+u-v}{h_2}$, we have
\begin{eqnarray}\label{eq_Un_2}
    E(U_n)&=&\intinfinf\intinfinf\intzeroinf\intzeroinf K_2(w) f_1(x)f_2(y)\notag\\
    &\times & f_{\epsilon_1}(v-m(x)+m(g_0(y))+wh_2)f_{\epsilon_2}(v)dxdydwdv.
\end{eqnarray}
From Assumption~1,~3 and~4, the integrand on the right hand side of
(\ref{eq_Un_2}) converges, as $n\rightarrow\infty$, to
\begin{eqnarray*}
K_2(w) f_1(x)f_2(y)
f_{\epsilon_1}(v-m(x)+m(g_0(y)))f_{\epsilon_2}(v),
\end{eqnarray*}
and is bounded by the integrable
function $$M_fK_2(w)f_1(x)f_2(y)
f_{\epsilon_2}(v).$$ By applying DCT, we have
\begin{eqnarray}\label{Limit_Un}
    \lim_{n\rightarrow\infty}E(U_n)&=&\intinfinf\intzeroinf\intzeroinf
     f_1(x)f_2(y)\notag\\
     &\times& f_{\epsilon_1}(v-m(x)+m(g_0(y)))f_{\epsilon_2}(v)dxdydv.
\end{eqnarray}

\noindent
We now turn to the variance of $U_n$. From (\ref{Un_exp}), we have
\begin{eqnarray*}
Var(U_n)&=& V_1+V_2+V_3+V_4,
\end{eqnarray*}
where
\begin{eqnarray}
\!\!\!\!\!\!V_1\!\!\!\!\!\!&=&\!\!\!\!\!\!\frac{1}{(n_1n_2h_2)^2}\SUM{i}{1}{n_1}\SUM{j}{1}{n_2}Var\left\{
K_2\left(\frac{ Y_{t_i}-Y'_{s_j}}{h_2}\right)\right\},\label{v(un)_v1}\\
\!\!\!\!\!\!V_2\!\!\!\!\!\!&=&\!\!\!\!\!\!\frac{1}{(n_1n_2h_2)^2}\SUM{i}{1}{n_1}\SUM{j}{1}{n_2}\nSUM{i'}{1}{n_1}{i}
Cov\left\{K_2\left(\frac{Y_{t_i}-Y'_{s_j}}{h_2}\right),K_2\left(\frac{Y_{t_{i'}}
-Y'_ {s_j}}{h_2}\right)\right\},\label{v(un)_v2}\\
\!\!\!\!\!\!V_3\!\!\!\!\!\!&=&\!\!\!\!\!\!\frac{1}{(n_1n_2h_2)^2}\SUM{i}{1}{n_1}\SUM{j}{1}{n_2}\nSUM{j'}{1}{n_2}{j}
Cov\left\{K_2\left(\frac{Y_{t_i}-Y'_{s_j}}{h_2}\right),K_2\left(\frac{Y_{t_i}
-Y'_ {
s_{j'}}}{h_2}\right)\right\},\\
\!\!\!\!\!\!V_4\!\!\!\!\!\!&=&\!\!\!\!\!\!\frac{1}{(n_1n_2h_2)^2}\SUM{i}{1}{n_1}\SUM{j}{1}{n_2}\nSUM{i'}{1}{n_1}{i}
\nSUM{j'}{1}{n_2}{j}Cov\left\{K_2\left(\frac{Y_{t_i}-Y'_{s_j}}{h_2}\right),
K_2\left(\frac{Y_{ t_{i'}}-Y'_{s_{j'}}}{h_2}\right)\right\}.
\end{eqnarray}

\noindent
We show that the terms $V_i$, for $i=1,\ldots,4$, converges to zero as $n\rightarrow\infty$. By making the transformation $w=\frac{m(x)-m(\g(y))+u-v}{h_2}$ from (\ref{v(un)_v1}) and ~(\ref{model1}), we have
\begin{eqnarray}\label{Un_V1}
V_1&=&\frac{1}{n_1n_2h_2}V_{11}-\frac{1}{n_1n_2}E^2(U_n),
\end{eqnarray}
where
\begin{eqnarray}\label{v(un)_v11}
    V_{11}&=&\intinfinf\intinfinf\intzeroinf\intzeroinf K_2^2(w)f_1(x)f_2(y)\notag\\
    &\times & f_{\epsilon_1}(v-m(x)+m(g_0(y))+h_2w)f_{\epsilon_2}(v)dxdydwdv.
\end{eqnarray}
From Assumption~1,~3 and~4, the integrand on the right hand side of
(\ref{v(un)_v11}) converges, as $n\rightarrow\infty$, to
\begin{eqnarray*}
K^2_2(w) f_1(x)f_2(y)
f_{\epsilon_1}(v-m(x)+m(g_0(y)))f_{\epsilon_2}(v)
\end{eqnarray*}
and is dominated by the integrable
function
$$M_f K^2_2(w)f_1(x)f_2(y) f_{\epsilon_2}(v).$$
Thus, by applying DCT, we have
\begin{eqnarray*}
    \lim_{n\rightarrow\infty}V_{11}&=&\intinfinf K^2_2(w) dw \\
    &\times& \intinfinf\intzeroinf\intzeroinf f_1(x)f_2(y)
    f_{\epsilon_1}(v-m(x)+m(g_0(y)))f_{\epsilon_2}(v)dxdydv.
\end{eqnarray*}
Now, from (\ref{Limit_Un}), the second term on the right hand side of (\ref{Un_V1}) turns out to be
\begin{eqnarray*}
\frac{1}{n_1n_2}E^2(U_n)=O\left(\frac1{n_1n_2}\right).
\end{eqnarray*}
Thus, we have
$$V_1=O\left(\frac{1}{n_1n_2h_2}\right) +
O\left(\frac{1}{n_1n_2}\right)=O\left(\frac{1}{n_1n_2h_2}\right).$$

\noindent
By making the transformations $w=\frac{m(x)-m(\g(y))+u-v}{h_2}$,
$w'=\frac{m(x')-m(g_0(y))+u'-v}{h_2}$ and using ~(\ref{model1}) and
~(\ref{v(un)_v2}), we have
\begin{equation}\label{Un_V2}
V_2=\frac{n_1-1}{n_1n_2}V_{21}-\frac{n_1-1}{n_1n_2}E^2(U_n),
\end{equation}
where
\begin{eqnarray}\label{v(un)_v21}
    V_{21}&=&\intinfinf\intinfinf\intinfinf\intzeroinf\intzeroinf\intzeroinf K_2(w)K_2(w')f_1(x)f_2(y)f_1(x')\notag\\
    &\times&f_{\epsilon_1}(v-m(x)+m(g_0(y))+h_2w)\notag\\
    &\times& f_{\epsilon_1}(v-m(x')+m(g_0(y))+h_2w')f_{\epsilon_2}(v)dxdydx'dwdw'dv.
\end{eqnarray}
From Assumption~1,~3 and~4, the integrand function on the right hand side of (\ref{v(un)_v21}) converges, as $n\rightarrow\infty$, to
\begin{eqnarray*}
    && \hskip -20pt K_2(w)K_2(w')f_1(x)f_2(y)f_1(x')
    f_{\epsilon_1} (v-m(x)+m(g_0(y)))\\
    &\times& f_{\epsilon_1}(v-m(x')+m(g_0(y)))f_{\epsilon_2}(v),
\end{eqnarray*}
and is dominated by the integrable function
$$M_f^2K_2(w)K_2(w')f_1(x)f_2(y)f_1(x')f_{\epsilon_2}(v).$$
Thus, by using DCT, we have
\begin{equation*}
\lim_{n\rightarrow\infty}V_{21}
=\intinfinf\intzeroinf
f_2(y)f_{\epsilon_2}(v)\left\{\intzeroinf
f_1(x)
 f_{ \epsilon_1 } (v-m(x)+m(g_0(y)))
dx\right\}^2dvdy.
\end{equation*}
Now, from (\ref{Limit_Un}), the second term on the right hand side of (\ref{Un_V2}) turns out to be
\begin{eqnarray*}
\frac{n_1-1}{n_1n_2}E^2(U_n)=O\left(\frac1{n_2}\right).
\end{eqnarray*}
Thus, we have
\begin{eqnarray*}
n_2V_2&\rightarrow& \intinfinf\intzeroinf
f_2(y)f_{\epsilon_2}(v)\left\{\intzeroinf
f_1(x)
 f_{ \epsilon_1 } (v-m(x)+m(g_0(y)))
dx\right\}^2dvdy \\
&& - \left\{\intinfinf\!\intzeroinf\!\!\intzeroinf\!
 f_1(x)f_2(y) f_{\epsilon_1}(v\!-\!m(x)\!+\!m(g_0(y)))f_{\epsilon_2}(v)
dxdydv\right\}^2\!\!,
\end{eqnarray*}
i.e.,
$$V_2=O\left(\frac1{n_2}\right).$$

A similar argument, shows that
\begin{eqnarray*}
 n_1V_3&\rightarrow& \intinfinf\intzeroinf
f_1(x)f_{\epsilon_1}(v)\left\{\intzeroinf
f_2(y)
 f_{ \epsilon_2} (v+m(x)-m(g_0(y)))
dy\right\}^2dvdx \\
&& - \left\{\intinfinf\!\intzeroinf\!\!\intzeroinf\!
 f_1(x)f_2(y) f_{\epsilon_1}(v\!-\!m(x)\!+\!m(g_0(y)))f_{\epsilon_2}(v)
dxdydv\right\}^2\!\!,
\end{eqnarray*}
i.e.,
$$V_3=O\left(\frac1{n_1}\right).$$ The term $V_4$ is seen to be $0$ from the model
specification. This completes the proof. $\hfill\Box$

\bigskip
\noindent
{\bf Proof of Theorem~\ref{m_thm3}:}
For any given $\epsilon > 0$, we have
\begin{eqnarray}
&&\hskip-20ptP\{|L_n(\hat{g}_n)-L(g_0)|>\epsilon\}\notag\\
&\leq&P\{|L_n(\hat{g}_n)-L(g_0)|>\epsilon,|L_n(g_0)-L(g_0)|\leq\epsilon\}
+P\{|L_n(g_0)-L(g_0)|>\epsilon\}\notag\\
&\leq& P\{|L_n(\hat{g}_n)-L(g_0)|>\epsilon,|L_n(g_0)-L(g_0)|\leq\epsilon,|L_n(\hat{g}_n)-L(\hat{g}_n)|\leq\epsilon\}\notag\\
&&+P\{|L_n(\hat{g}_n)-L(\hat{g}_n)|>\epsilon\}
+P\{|L_n(g_0)-L(g_0)|>\epsilon\},\label{thm3_2}
\end{eqnarray}
where $\hat{g}_n$ is as in (\ref{cr2}). We will complete the proof by
establishing that all the three terms on the right
hand side of (\ref{thm3_2}) are arbitrarily small.

\noindent
We begin with the first term on the right hand side of (\ref{thm3_2}).
Note that, from (\ref{cr2}), we have $$L_n(g_0)\le L_n(\gn).$$ Therefore, from (\ref{cr2}), we have
\begin{eqnarray}\label{thm3_3}
\mbox{if}~|L_n(g_0)-L(g_0)|\leq\epsilon\quad\mbox{then}\quad  L_n(\hat{g}_n)\ge
L(g_0)-\epsilon.
\end{eqnarray}
We now turn to computing an upper bound for $L_n(\hat{g}_n)$ in terms of
$L(g_0)$. From Theorem~\ref{m_thm1.5}, we have $L(\hat{g}_n)\le L(g_0)$. Therefore,
\begin{eqnarray}\label{thm3_4}
\mbox{if}~|L_n(\hat{g}_n)-L(\hat{g}_n)|\leq\epsilon\quad\mbox{then}\quad
L_n(\hat{g}_n)\le
L(g_0)+\epsilon.
\end{eqnarray}
Further, (\ref{thm3_3}) and (\ref{thm3_4}) imply that
\begin{eqnarray}\label{thm3_6}
    \mbox{if}~
    |L_n(g_0)-L(g_0)|\leq\epsilon\;\mbox{and}\;|L_n(\hat{g}_n)-L(\hat{g}
    _n)|\leq\epsilon\;\mbox{then}\,
    |L_n(\hat{g}_n)-L(g_0)|\leq\epsilon.
\end{eqnarray}
Thus, from (\ref{thm3_6}),
$$P\{|L_n(\hat{g}_n)-L(g_0)|>\epsilon,|L_n(g_0)-L(g_0)|\leq\epsilon,|L_n(\hat{g}
_n)-L(\hat{g}_n)|\leq\epsilon\}=0,$$
which takes care of the first term on the right
hand side of (\ref{thm3_2}).

\noindent
We now consider the second term.
Observe that
\begin{eqnarray}\label{thm3_7}
 |L_n(\gn)-L(\gn)|\leq\sup_{g \in\G}|L_n(g)-L(g)|.
\end{eqnarray}
From (\ref{thm3_7}) and Theorem~\ref{m_thm2}, we have
$$L_n(\gn)-L(\gn){\stackrel{P}{\longrightarrow}}0.$$
This ensures that the second term on the right hand side of (\ref{thm3_2}) goes
to zero as $n\rightarrow\infty$. Further, Theorem~(\ref{m_thm1}) ensures that
the last term on
the right hand side of  (\ref{thm3_2}) goes to zero too.
This completes the proof. $\hfill \Box$

\bigskip
\noindent
{\bf Proof of Theorem~\ref{m_thm4}:}
If $\hat{g}_n\nconv{P}g_0$, then there exists an $\epsilon>0$ and a $\delta
>0$ such that
\begin{eqnarray}\label{thm4_0}
P\{\sup|\hat{g}_n-g_0|\geq\epsilon\}>\delta\;\mbox{infinitely often}.
\end{eqnarray}

\noindent
Note that,
$\mathcal{N}_\epsilon^c(g_0)=\{g:\sup|g-g_0|\ge\epsilon,\;g\in\G\}$ is a closed
subset of
$\G$. From Assumption~5 and Lemma~\ref{lem_UC}, there exists
a
$\tilde{g}\in \mathcal{N}_\epsilon^c(g_0)$ such that
$\tilde{g}=\argmax_{g\in\mathcal{N}_\epsilon^c(g_0)}L(g)$. It follows from
part~(b) of Theorem~\ref{m_thm1.5} that the supremum of the
functional $L$ is attained only at
$g_0$. Therefore, $\hat{g}_n\in\mathcal{N}_\epsilon^c(g_0)$ implies
\begin{equation}\label{thm4_1}
|L(g_0)-L(\hat{g}_n)|=L(g_0)-L(\hat{g}_n)\ge L(g_0)-L(\tilde{g})>0.
\end{equation}
Denote $\eta=L(g_0)-L(\tilde{g})$. By using the triangular inequality, we have
\begin{eqnarray}\label{thm4_2}
  |L(g_0)
-L_n(\hat{g}_n)|+|L_n(\hat{g}_n)-L(\hat{g}_n)|&\ge& |L(g_0)-L(\hat{g}_n)|.
\end{eqnarray}
From (\ref{thm4_1}) and (\ref{thm4_2}),
$\hat{g}_n\in\mathcal{N}_\epsilon^c(g_0)$ implies
\begin{eqnarray}\label{thm4_3}
 |L(g_0)
-L_n(\hat{g}_n)|&\ge& \eta-|L_n(\hat{g}_n)-L(\hat{g}_n)|.
\end{eqnarray}
Now from (\ref{thm4_3}),
\begin{eqnarray}\label{thm4_4}
\mbox{if}~\hat{g}_n\in\mathcal{N}_\epsilon^c(g_0)~\mbox{then}~
\sup_{g\in\mathbb{G}}|L_n(g)-L(g)|<\frac{\eta}{2}~\mbox{implies}~
|L_n(\hat{g}_n)-L(g_0)|>\frac{\eta}{2}.
\end{eqnarray}
Therefore, from (\ref{thm4_4}), we have
\begin{eqnarray}\label{thm4_5}
    && \hskip -20pt P\bigg\{|L_n(\hat{g}_n)-L(g_0)|>\frac{\eta}{2}\bigg\}\notag\\
    &\geq& P\bigg\{\hat{g}_n\in\mathcal{N}_\epsilon^c(g_0)~\mbox{and}~\sup_{g\in\mathbb{G}}
    |L_n(g)-L(g)|<\frac{\eta}{2}\bigg\},\notag\\
    &\geq&P\{\sup|\hat{g}_n-g_0|\geq\epsilon\}+P\bigg\{\sup_{g\in\mathbb{G}}
    |L_n(g)-L(g)|<\frac{\eta}{2}\bigg\}-1.
\end{eqnarray}
From (\ref{thm4_0}), the first term on the right hand side of (\ref{thm4_5}) is
greater than $\delta$ infinitely often.
From Theorem~\ref{m_thm2}, the second term on the right hand side of
(\ref{thm4_5}) is
greater than $1-\frac{\delta}{2}$ for all but finitely many $n$.
Therefore,
\begin{eqnarray*}
P\left\{|L_n(\hat{g}_n)-L(g_0)|>\frac{\eta}{2}\right\}&>&\frac\delta2\quad\mbox{
infinitely often}.
\end{eqnarray*}
This contradicts Theorem~\ref{m_thm3} and completes the proof.$\hfill
\Box$

\newpage

\begin{table}[ht!]
    \begin{center}

        \caption {Average normalized IMSE (in $10^{-3}$)}\label{tab:imse}
        \begin{tabular}{lrrrr}
            \hline
            Method&Scen.\ 1&Scen.\ 2&Scen.\ 3&Scen.\ 4\\
            \hline
            Cont.\ mon.\ registration&1.000&1.022&0.598&0.108\\
            Self-modelling registration&0.375&--&0.433&--\\
            The proposed method&0.104&0.287&0.179&0.189\\
            \hline
        \end{tabular}
        \newline
        \newline
        \caption{Some descriptive statistics}\label{tab:des_stat}
        \begin{tabular}{lrcrc}
          \hline
           & \multicolumn{2}{c}{Data set 1: Vostok} & \multicolumn{2}{c}{Data set 2: EPICA
    Dome} \\
          \cline{2-5}
          Data&Size&Range($Y$-Value)&Size&Range($Y$-Value)\\
         \hline
         Carbon dioxide&283&182.2-298.7&537&183.8-298.6\\
         Methane&457&318-773&1545&342-907\\
         Temp. deviations&3,310&(-)9.39-3.23&5028&(-)10.58-5.46\\
          \hline
        \end{tabular}
        \newline
        \newline
        \caption{Average squared difference between pair of data sets}\label{tab:sq_diff}
        \begin{tabular}{lrrr} \hline
            &Carbon dioxide&Methane&Temp.\ dev.\ \\ \hline
            Pre-alignment&266.09&3739.06&3.04\\ \hline
            Post-alignment:&&&\\
                Cont.\ mon.\ registration&70.76&--&1.76\\
                Proposed method&16.20&1353.93&1.31\\
            \hline
        \end{tabular}
    \end{center}
\end{table}

\end{document}


\thispagestyle{empty}
\begin{center}
 \textbf{\large SUPPLEMENTAL MATERIALS}\\
\end{center}

\begin{figure}[h!]
 \centering
  \includegraphics{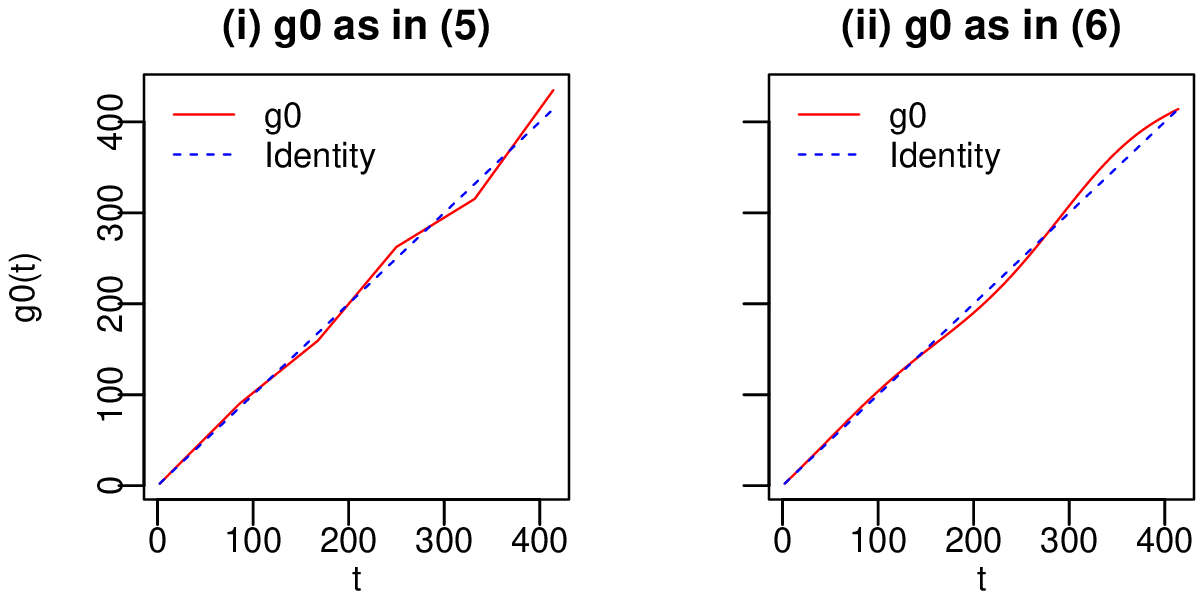}
  \caption{$g_0$ functions for simulations}\label{fig:sim_g0}
\end{figure}

\begin{figure}[h!]
 \centering
  \includegraphics{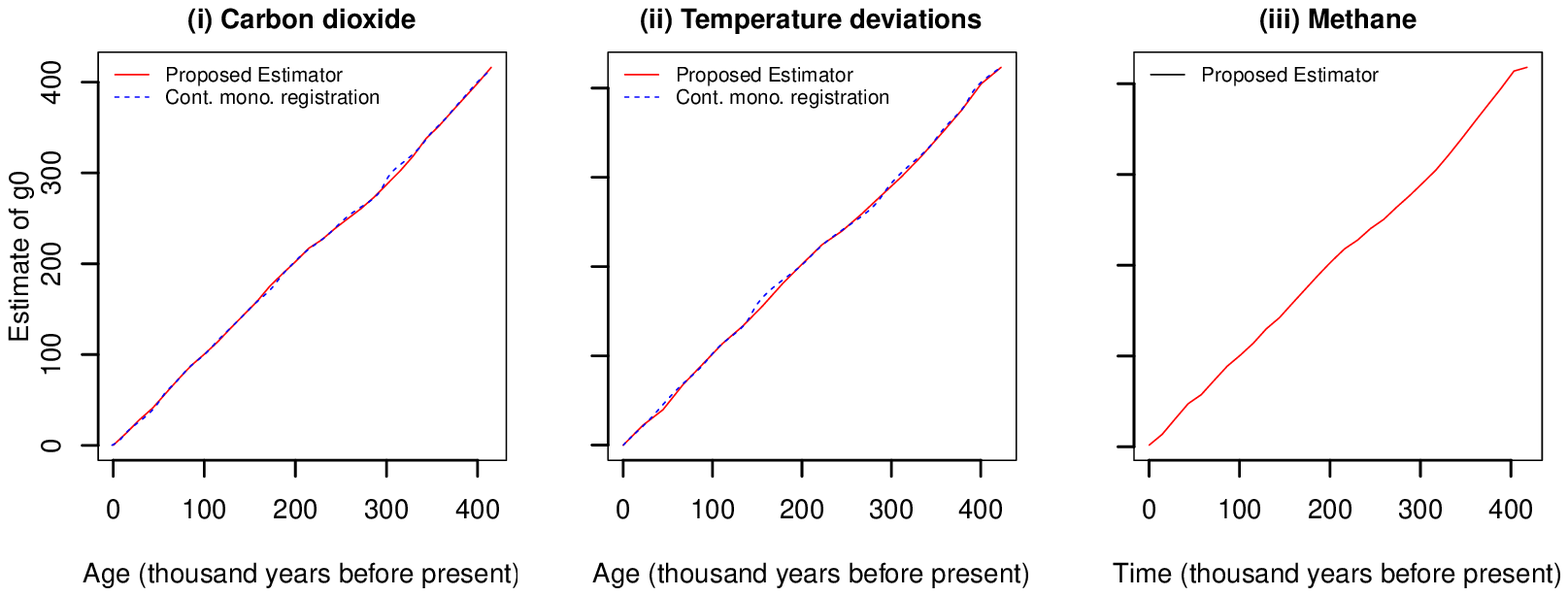}
  \caption{Estimates of $g_0$ for paleoclimatic data sets}\label{fig:est_g0}
\end{figure}